\documentclass[12pt]{article}

\usepackage[margin=0.75in]{geometry}

\usepackage{amssymb}
\RequirePackage{booktabs}
\usepackage{multirow}
\usepackage{ulem}
\usepackage{xcolor}
\usepackage{changepage}

\usepackage{amsmath}
\usepackage{graphicx}
\usepackage[colorlinks=true, allcolors=blue]{hyperref}
\usepackage{fontenc}
\usepackage{prodint}
\usepackage{natbib}

\newtheorem{thm}{Theorem}
 \newtheorem{remark}{Remark}
 
\usepackage{comment,cleveref}

\title{R-Estimation with Right-Censored Data}
\author{Glen A. Satten \\
Department of Gynecology and Obstetrics \\
School of Medicine, Emory University, Atlanta GA
U.S.A. \\[.5ex]
Mo Li \\
Department of Mathematics \\
University of Louisiana at Lafayette, Lafayette LA
U.S.A. \\[.5ex]
Ni Zhao \\
Department of Biostatistics \\
Johns Hopkins University, Baltimore MD
U.S.A.\\[.5ex]
Robert L. Strawderman 
\footnote{Corresponding author: \href{robert\_strawderman@urmc.rochester.edu}{robert\_strawderman@urmc.rochester.edu}}  
\\ Department of Biostatistics and Computational Biology 
\\ University of Rochester, Rochester NY U.S.A. 
}

\date{}
\begin{document}
\maketitle


\begin{abstract}
This paper considers the problem of directly generalizing the R-estimator under a linear model formulation with right-censored outcomes. We propose a natural generalization of the rank and corresponding estimating equation for the R-estimator in the case of the Wilcoxon
(i.e., linear-in-ranks) score function, and show how it can respectively be exactly represented as members of the classes of estimating equations proposed in \citet{ritov1990} and \citet{tsiatis1990}. We then establish analogous results for a large class of bounded nonlinear-in-ranks score functions. Asymptotics and variance estimation are obtained as straightforward consequences of these representation results. The self-consistent estimator of the residual distribution function, and the mid-cumulative distribution function (and, where needed, a generalization of it), 
play critical roles in these developments. \\[.5ex]
\end{abstract}

\newpage
\setcounter{page}{1} 

\section{Introduction }
The linear model is the fundamental workhorse for much of statistics.  Although ordinary least squares is the most common way to fit the linear model, it is known that least-squares estimators are sensitive to outlying observations.  For this reason, a variety of robust estimators for estimating the parameters in the linear model have been proposed, including but not limited to R-estimators. Because R-estimators are rank-based, they are more robust to the shape of the underlying error (i.e., residual) distribution. Some influential early work in this area includes \citet{sen1968}, \citet{JJ1972} and \citet{jaeckel1972estimating}; see, for example, \citet{Hettmansperger84,chang1999, Hettmansperger2e}
for selected later developments.

A well-known exception to the popularity of the linear model is the Cox proportional hazards model, which has long been the workhorse of survival analysis.  The Accelerated Failure Time (AFT) model \citep[e.g.,][Ch.\ 7]{KalbfleischPrentice2e} is a direct analog of the linear model, and in its most common form, models the natural log of the failure time as a linear function of available covariates. The AFT model can easily be fit using maximum likelihood when one imposes parametric assumptions on the ``error'' distribution
\citep[e.g.,][Sec.\ 6.5]{lawless82}; computations and inference become considerably more challenging in the semiparametric setting, where the error distribution is left unspecified. The AFT model is favored when the parameters in the log-linear model are of direct interest, or otherwise required \citep[e.g.,][]{Satten2025}.

In a linear model, the regression function is interpretable as the (conditional) mean in the case where the error distribution is assumed to have mean zero and, as is frequently assumed, symmetric about the mean. In this context, R-estimators remain an attractive choice for parameter estimation due to their reduced sensitivity to skewness and outliers \citep{Hettmansperger2e}. In the case of the AFT model, the regression function is not typically interpreted as a model for a mean function. This is because the ``error'' distribution, typically allowed to have a non-zero mean, intends to capture baseline survivorship experience for the subpopulation that is defined by setting all covariates to zero. Correspondingly, this underlying distribution may generate ``errors'' that are asymmetrically distributed about its mean, and the reduced sensitivity of R-estimators to such behavior is an attractive feature
around which to build estimators.

There are numerous proposals for estimators of the AFT regression model parameters with right-censored data. Highly influential early works in the case of the semiparametric AFT model include those of \citet{prentice1978} and \citet{tsiatis1990}. \citet{prentice1978} developed the theory of linear rank tests for right-censored data in the context of testing hypotheses for the regression coefficient under the AFT model, and suggested estimating the same by turning the proposed score test statistic into 
an objective function; see \citet[][Sec.\ 8.2]{lawless82} and \citet[][Ch.\ 7]{KalbfleischPrentice2e} for helpful reviews.
\citet{tsiatis1990} built on this idea, and established the relevant
asymptotic theory for the corresponding estimator; see also \citet{MMM82},
\citet{cuzick85}, and \citet{ritov1990} for related developments that connect the two approaches. Estimation is complicated in the context of the semiparametric model due to an inherent lack of smoothness. The work of \citet{FygensonRitov94} is notable in this regard, as they develop the ``Gehan estimator'' by developing a valid estimation procedure that involves minimizing a convex objective function. Since then, significant effort has been devoted to improving the computational aspects of estimating AFT regression model parameters and associated methods of inference and variance estimation; two (of many) examples here include include those \citet{JinLin03} and
\citet{johnstraw09}.

The Gehan estimator has some very nice properties due to its derivation as a minimizer of
a convex objective function. However, considered as an  member of the broader class of weighted logrank estimators, it is problematic from an asymptotic efficiency perspective. In particular, it is known that the most efficient choice of weight function within this class depends only on the underlying failure distribution \citep[e.g.,][]{tsiatis1990}; in contrast, the choice of weight function within the class of 
weighted logrank estimators that leads to the Gehan estimator necessarily depends on the underlying censoring distribution. Moreover, in line with the particular interests of this paper, the Gehan estimator is not truly a generalization of the R-estimator because it makes use of `rankings' only among those observations that can be unambiguously ordered. 

In this paper, our goal is to develop an estimating equation for the AFT model in the presence of right-censored data that follows the development of the corresponding full-data R-estimator as closely as possible.  Motivated by the notion of censoring-unbiased transformations, we achieve this goal by replacing the actual ranks with an imputed ranking that shares similar behavior.  The so-called mid-cumulative distribution function
(CDF) plays an important role in this development in the case of estimating functions that use linear transformations of ranks; in the case of nonlinear transformations, a generalization of the mid-CDF is required and introduced. The remainder of this paper is organized as follows.  After introducing some notation in Section \ref{sec:notation}, we develop our proposed generalization of the R-estimator to right-censored data in Section \ref{sec:R-est}, focusing on the case of the Wilcoxon score (i.e., a linear transformation). In Section \ref{sec:genR}, we show that this proposed generalization  has direct connections to two other important classes of rank-based estimators.  In Section \ref{sec:variance}, we use these results to derive a variance estimator for our parameter estimates. In Section \ref{sec:nonlinear score} we extend our approach to allow for the use of nonlinear transformations of ranks.  We study the performance of our estimators through simulation in Section \ref{sec:sim}.  Finally, we conclude the paper in Section \ref{sec:discuss} with a discussion.

\section{Notation}
\label{sec:notation}

We assume the expected value of outcome $Y^*_i$ is a linear function of a 
$p-$vector of covariates $X_{i},$ specified through the linear model
\begin{equation}
\label{linear.model}
    Y^*_i= X_{i}^T \beta_0  + \epsilon_i.
\end{equation}
In this model, the ``errors'' $\epsilon_i$ are assumed to have a common unknown CDF $F$. A more typical formulation of \eqref{linear.model} might include $\alpha$ as an intercept term and assume $E[\epsilon_i | X_i] = 0;$ we intentionally avoid introducing an intercept in \eqref{linear.model}, instead treating it as a nuisance parameter for the remainder of this paper. As such, $\beta$ does not contain an intercept parameter, and we have $E[\epsilon_i | X_i] = \int u d F(u) = \alpha,$ where $\alpha \in \mathbb{R}$ is assumed finite. 

The accelerated failure time model (AFT) corresponds to \eqref{linear.model} when $Y^*_i=\log(T^*_i),$ where $T^*_i > 0$ denotes a failure time. In this case, the conditional survivor function $P\{ T^*_i > u | X_i\}  = F(\log u - X^T_i \beta_0),$ where $\epsilon_i \sim F.$  Failure time data are often right-censored, in which case we also posit the existence of a potential right-censoring time $C^*_i,$ and where it is assumed that one can only observe $T_i=\min(T^*_i,C^*_i)$, $Y_i=\log(T_i)$ and $\Delta_i=I\{C^*_i \ge T^*_i\}$.  We impose the standard identification condition $T^*_i \perp C^*_i | X_{i}$ for each $i$ \citep{KalbfleischPrentice2e}. For the moment, we do not impose any continuity assumptions on distributions of $T^*$ and $C^*.$

\section{R-Estimation and Inference}
\subsection{Brief review: complete data setting}
\label{sec:U-est}
With uncensored data, the regression coefficient $\beta_0$ in \eqref{linear.model} can be estimated by the so-called R-estimator $\widehat{\beta}$, defined as any minimizer of the objective function
\begin{equation*}
    \sum_{i=1}^n a( R(\tilde E_{i,\beta}) )  \tilde E_{i,\beta}
\end{equation*}
where $a: [0,1] \rightarrow \mathbb{R}$ is a monotone increasing score function and $n R(\tilde E_{i,\beta})$ denotes the rank of the $i^\text{th}$ ``residual''
\begin{equation}
\label{eq:Etil}
    \tilde E_{i,\beta}=Y_i - X_{i}^T \beta, 
\end{equation}
among all $n$ $\beta-$dependent residuals \citep[e.g.,][]{jaeckel1972estimating}.  Equivalently, one can find a solution to
\begin{equation}
\label{R.est}
\sum_{i=1}^n a( R(\tilde E_{i,\beta}) ) \left( X_i - \bar X \right)  = 0
\end{equation}
where $\bar X = n^{-1} \sum_{i=1}^n X_i$; see, for example, \citet{Hettmansperger84, Hettmansperger2e}. Importantly,
the solution to \eqref{R.est} is invariant under any location shift; hence, the fact that $\epsilon_i$ does not necessarily 
have mean zero under model \eqref{linear.model} for any $i$ does not affect its validity.

\citet{JJ1972} establishes the asymptotic linearity, hence basis for inference, of \eqref{R.est} in $\beta$ for general score functions $a(\cdot)$ and \citet{jaeckel1972estimating} establishes the corresponding asymptotic distribution of $\widehat \beta;$ see \citet[Thm.\ 5.2.3]{Hettmansperger84} for a summary of these results, where a detailed analysis for the score $a(u) = u$ (i.e., the ``Wilcoxon'' score function, defined up to a known location and scale) is also provided. In this case,  one simple but natural estimator for $F$ is the empirical CDF of the estimated residuals $\tilde E_{i,\widehat \beta}, i = 1,\ldots,n.$ Correspondingly, when $F$ is additionally assumed to be symmetric, the median of $\widehat F$ can be used as a simple
estimator of the unknown intercept term $\alpha,$ if desired; in fact, even when $F$ is not assumed to be symmetric, the median of $\widehat F$ remains a possible estimator of location \citep[Sec.\ 5.2]{Hettmansperger84}.

\subsection{Extending R-Estimation to Right-Censored Data}
\label{sec:R-est}

\subsubsection{Estimating $\beta_0$: the case
of the Wilcoxon score}
\label{sec:censored-Wilcoxon}
As noted in the introduction, numerous rank-based estimators for the AFT model with right-censored data have been proposed; critical papers in this area include \citet{tsiatis1990} and \citet{FygensonRitov94}
and important related work includes, but is not limited to, \citet{prentice1978}, \citet{cuzick85},
\citet{ritov1990} and \citet{JinLin03}. The central problem for rank-based methods when data are right-censored is that 
the ranks of censored observations cannot be fully observed. Typically, existing methods can be 
distinguished by how each addresses this problem.

\citet{sen1968} suggested an estimator for $\beta$ under \eqref{linear.model} by considering
a modification of Kendall's $\tau$ statistic. \citet{FygensonRitov94} directly generalize this
approach to estimation for the AFT model, resulting in a monotone estimating function that
only involves the contributions of pairs of residuals that can be unambiguously ordered (i.e., ranked).
If we define
\begin{equation}\label{FR-ranks}
    \mathcal{D}_{i,j}(\beta)
    = \Delta_i I\{ \tilde E_{i,\beta} < \tilde E_{j,\beta}  \},
\end{equation}
where $\tilde E_{i,\beta}$ is defined as in \eqref{eq:Etil}, then it can be shown that the Fygenson and Ritov estimator is equivalent to solving 
\begin{equation}\label{FR-eqn0}
    \sum_{i=1}^n 
    \left( \mathcal{D}_{i, +}(\beta)
    - \mathcal{D}_{+, i}(\beta)
    \right)
 \left(X_i-\bar X\right)=0,
\end{equation}
where the `+` subscript denotes summation over that index. In the absence of censoring and ties between
residuals, the estimating function \eqref{FR-eqn0} involves considering an $X-$weighted contrast between the rank and
anti-rank of each residual, and it can be shown that this reduces to \eqref{R.est} for the Wilcoxon score 
$a(u) = u$ (i.e., up to centering and scaling constants). However, in the presence of right-censoring, this estimating function only involves all pairwise rankings of uncensored residuals, and does not use an obvious analog to the notion of ``rank'' with a right-censored sample. 

Here, we take an alternative approach and attempt to directly generalize \eqref{R.est} by assigning a rank 
for each observation that \textit{is} directly related to the rank it would have in the uncensored experiment.  In this section, we will focus on the Wilcoxon score, or $a(u)=u$; extensions to more general rank transformations will be considered in Section \ref{sec:nonlinear score}.  Our motivation is  the fact that, for uncensored continuous data, the rank of the $i^{th}$ ``residual'' $\epsilon_i$ equals $n \widehat{F}(\epsilon_i),$
where $\widehat{F}$ denotes the empirical distribution of the $\epsilon_i$s. In the case of a right-censored sample, ranks can be analogously defined for uncensored observations upon replacing the empirical distribution $\widehat{F}$ by an estimator that appropriately handles censored data, such as the Kaplan-Meier estimator or the self-consistent estimator \citep[e.g., see ][]{Strawderman2023}. However, this same definition cannot be used for censored observations. Instead, considering that the population analog of $n \widehat{F}(\epsilon_i)$ is $n F(\epsilon_i),$
an obvious choice is to use $E[ n F(\epsilon_i) | \epsilon_i > u]$ for any observation right-censored at time $u,$ which corresponds to averaging the unobserved rank over all ``residuals'' exceeding $u$.

For reasons that will become clear later, it will be helpful to use mid-ranks in our development.  In particular, for the CDF $F$ of the residuals, let the corresponding mid-CDF be  
\begin{equation}\label{mid-CDF}
    F^{\star}(y)=\frac{1}{2} \left[ F(y)+F(y-) \right];
\end{equation}
similarly, given a suitable estimator $\widehat F,$ let $\widehat{F}^{\star}$ be the corresponding plug-in estimator derived from \eqref{mid-CDF}. In the absence of censoring and in the possible presence of ties, a mid-rank can be defined as $n \widehat F^{\star}(\epsilon_i)$ \citep{ruym80}. Following the proposed approach for censored data in the previous paragraph and for a given $\beta$, we shall define the imputed rank (or imputed mid-rank) for right-censored data as $n \widehat{\mathcal{R}} ( \tilde E_{i,\beta} ),$ where
\begin{eqnarray}
\label{est.rank.bj}
     \widehat{\mathcal{R}} ( \tilde E_{i,\beta} ) & = & 
     \Delta_i  \widehat{F}^{\star}( \tilde E_{i,\beta} ) + (1-\Delta_i)\int_{ \tilde{E}_{i,\beta}}^\infty \widehat{F}^{\star}(y) \frac{d\widehat{F}(y)}{\widehat{S}(\widetilde{E}_{i,\beta})} \: \:,
\end{eqnarray}
and $\widehat{S}=1-\widehat{F}$ is the corresponding estimated survival function. 

For the remainder of this paper, we will
assume that $\widehat S$ is the self-consistent estimator for $S$ derived from $(\Delta_i,\tilde E_{i,\beta}), i=1,\ldots,n$ \citep[e.g.,][]{Strawderman2023}. The self-consistent estimator is equivalent to the Kaplan-Meier estimator for $t < T_{(n)}$
and is zero for $t \geq T_{(n)}$ \citep[e.g.,][Prop.\ 2]{Strawderman2023}; equivalently, it is the Kaplan-Meier estimator calculated
in a modified dataset for which all times
$T_i = T_{(n)}$ have $\Delta_i = 1$ (i.e., regardless of their actual failure status). Going forward, we will assume that this modified
dataset is used in all empirical expressions; doing so will preserve compatibility with the use of $\widehat F$ and quantities that 
can be derived from it.  Generally, the estimator $\widehat{S},$ hence $\widehat F$ and $\widehat F^{\star},$ also depend on $\beta;$ 
for the remainder of this subsection, we suppress this dependence only for notational convenience. 

One immediate and useful consequence of replacing $\widehat F$ by $\widehat F^{\star}$ in \eqref{est.rank.bj} is that
\begin{equation}
\label{est.rank.2}
    \int_{ \tilde{E}_{i,\beta}}^\infty \widehat{F}^{\star}(y) \frac{d\widehat{F}(y)}{\widehat{S}(\widetilde{E}_{i,\beta})} =\frac{1}{2} \frac{\widehat{F}^2(\infty)-\widehat{F}^2( \tilde E_{i,\beta})}{\widehat{S} (\tilde E_{i,\beta})}=\frac{1}{2}\left[ 1+\widehat{F}( \tilde E_{i,\beta})    \right],
\end{equation}
where the second equality holds because
$\widehat{F}(\infty)=1$ for the self-consistent estimator; see the Appendix for further details.
Hence, \eqref{est.rank.bj} reduces to  the easy-to-compute quantity
\begin{eqnarray}
\label{est.rank.3}
     \widehat{\mathcal{R}} ( \tilde E_{i,\beta} ) 
     & = & \Delta_i  \widehat{F}^{\star}( \tilde E_{i,\beta} ) + (1-\Delta_i)\left[ 1-\frac{1}{2} \widehat{S}( \tilde E_{i,\beta} ) \right].
\end{eqnarray}
We stress that $n \widehat{\mathcal{R}} ( \tilde E_{i,\beta})$ involves both the self-consistent estimator $\widehat S(\cdot)$ and its counterpart calculated using the corresponding mid-CDF $\widehat F^{\star}(\cdot),$ where $\widehat F = 1 - \widehat S.$ Importantly, these results do not
hold as stated if one replaces $\widehat{F}^{\star}(\cdot)$ with $\widehat{F}(\cdot),$ highlighting the important role of the mid-CDF in these
two identities. The following result provides justification, and additional insight, into the population-level behavior of \eqref{est.rank.bj}, equivalently, \eqref{est.rank.3}.
\begin{thm}
\label{nice result}
Define $F$ to be the CDF of $\epsilon,$ and let
\[
\mathcal{R}_{\beta_0} = \Delta F^\star(\tilde E_{\beta_0}) + (1-\Delta) \frac{1}{S(\tilde E_{\beta_0})} \int_{\tilde E_{\beta_0}}^{\infty} F^\star(u) d F(u),
\]
where $\tilde E_{\beta_0} = Y - X^T \beta_0$ and $\Delta$ are respectively the true observed residual and failure status. Then, 
\begin{equation}
\label{eq:simple-R}
\mathcal{R}_{\beta_0}  
= \Delta (1- S^\star(\tilde E_{\beta_0})) + (1-\Delta) \left[ 1 - \frac{1}{2} S(\tilde E_{\beta_0})\right],
\end{equation}
and $E(\mathcal{R}_{\beta_0} \mid X) = E(\mathcal{R}_{\beta_0}) = 1/2.$
In the case where $F$ is continuous,
it is additionally true that
\[
\mbox{Var}(\mathcal{R}_{\beta_0} | X) = \frac{1}{12} \left( 1 - E( S^3(E_{\beta_0}) | X) \right),
\] 
where 
$E_{\beta_0} = \log C^* - X^T \beta_0,$ in which case
$\mbox{Var}(\mathcal{R}_{\beta_0}) = \frac{1}{12} \left( 1 - E( S^3(E_{\beta_0}) \right).$
\end{thm}

For a set of uncensored residuals and $F$ continuous, $F(E_{\beta_0,i})$ is uniformly distributed
on $(0,1).$ Hence, one implication of  Theorem \ref{nice result} is that the first two moments of $\mathcal{R}_{\beta_0,i}$ 
are very close to that of a uniform random variable as long as censoring is not too heavy. We next show that, when the empirical mid-CDF $\widehat F^{\star}$ is used to define \eqref{est.rank.3}, the sum of imputed ranks is a constant that is independent of $\beta$.

\begin{thm}
\label{nice result 2}
When the self-consistent estimator
of $F$ is used and $\| \beta \| < \infty,$
\begin{equation}
\label{est.ranksum}
\sum_{i=1}^n \widehat{\mathcal{R}} ( \tilde E_{i,\beta} )
= \frac{n}{2}.
\end{equation}
\end{thm}

Theorem \ref{nice result 2} shows that the sum of the mid-ranks, or $\sum_{i=1}^n n \widehat{\mathcal{R}} ( \tilde E_{i,\beta} ),$
is $n^2/2$ regardless of $\beta.$ The proof of this result is a special case of Theorem \ref{really nice result}, given later for the case of a nonlinear $a(\cdot),$ and is therefore deferred. Again, this result does not hold as stated if one replaces $\widehat{F}^{\star}(\cdot)$ with $\widehat{F}(\cdot)$ in \eqref{est.rank.3}. Considering $\eqref{R.est}$ and the developments above, we can thus generalize \eqref{R.est} by ``solving'' the equation $\Psi_n(\beta) = 0,$ where
\begin{equation}\label{R.est.cens}
    \Psi_n(\beta) = \sum_{i=1}^n  \widehat{\mathcal{R}} ( \tilde E_{i,\beta} ) ( X_{i} - \bar X);
\end{equation}
equivalently, due to Theorem \ref{nice result 2}, we can use
\[
\Psi^{(c)}_n(\beta) = \sum_{i=1}^n   \Bigl(  \widehat{\mathcal{R}}( \tilde E_{i,\beta} )- \frac{1}{2} \Bigr)  X_{i}
\]
in place of $\Psi_n(\beta).$ An exact solution may not exist due to the discontinuous nature of $\Psi_n(\beta),$
in which case one can define a solution $\hat \beta$ as any zero crossing of $\Psi_n(\beta).$
The resulting estimator directly generalizes the  R-estimator that solves $\eqref{R.est}$ for the Wilcoxon score 
$a(u) = u$ (or, equivalently, the mean zero score $a(u) = u -1/2$) to the case of right-censored data, and provides 
an interesting alternative to Fygenson and Ritov estimator that solves \eqref{FR-eqn0}. Further comments
on estimating $\beta$ in practice are given in Section \ref{sec:variance}.

\subsubsection{Connections between \eqref{R.est.cens}
and the Ritov and Tsiatis classes}
\label{sec:genR}

Let $\widehat S_\beta(\cdot)$ be the self-consistent estimator of $S$ derived from $(\Delta_i,\tilde E_{i,\beta}), i=1,\ldots,n.$ Under model \eqref{linear.model} with $Y$ possibly being right-censored, Ritov (\citeyear{ritov1990}) proposed to estimate $\beta$ by (approximately) solving ${\cal S}_{n,R}(\beta,\gamma) = 0,$ where
\begin{equation}
\label{G-fun}
{\cal S}_{n,R}(\beta,\gamma) = 
\sum_{i=1}^n
(X_i - \bar X) \left[ \Delta_i  \gamma(\tilde E_{i,\beta}) + (1-\Delta_i) \frac{1}{\widehat S_\beta(\tilde E_{i,\beta})} \int_{\tilde E_{i,\beta}}^\infty  \gamma(u) d \widehat F_{\beta}(u) \right]
\end{equation}
and where the given function $\gamma(\cdot)$ satisfies certain regularity conditions. The selection $\gamma(u) = u$ gives the Buckley-James estimator; the optimal choice of this function, in the case where $\epsilon_i$ has nice smooth distribution function, is obtained under a maximum likelihood framework and equals $\gamma_{opt}(u) = \frac{d}{du} \log f(u),$ where $f$ is the probability density function (PDF) corresponding to $F$. Setting $\gamma(u) = F(u)$ is optimal when $F$ is the logistic distribution.  

\citet{ritov1990} showed that any estimating function in the class \eqref{G-fun} is equivalent to a member of the class of weighted logrank (or linear rank) estimating equations introduced in Tsiatis (\citeyear{tsiatis1990}); related results are established in \citet{cuzick85}, and for the case of linear rank tests, by \citet{MMM82}.
The Tsiatis class of estimating equations has the form 
\begin{equation}
\label{T-fun}
{\cal S}_{n,T}(\beta; \gamma)  = \sum_{i=1}^n \int_{-\infty}^{\infty} w(u; \beta) ( X_i - \bar{X}(u; \beta) ) d N_i(u;\beta),
\end{equation}
where $N_i(u;\beta) = \Delta_i I\{ \tilde E_{i,\beta} \leq u\},$ and for $Y_i(u;\beta) = I\{ \tilde E_{i,\beta} \geq u\},$  
\[
\bar{X}(u; \beta) = \frac{\sum_{j=1}^n X_j Y_j(u;\beta)}{\sum_{j=1}^n Y_j(u;\beta)}.
\]
\citet{ritov1990} (see also \cite{ABGK}, pp 582-583) establishes a direct connection between \eqref{G-fun} and \eqref{T-fun}; in particular, mathematical equivalence occurs when 
\begin{equation}
\label{equiv-wt}
w(u; \beta) = \gamma(u) - \frac{1}{\widehat S_\beta(u)} \int_{u}^\infty \gamma(u) \, d \widehat F_{\beta}(u).
\end{equation}

Comparing \eqref{R.est.cens} and \eqref{G-fun}, it is immediately evident from \eqref{est.rank.bj} 
that setting $\gamma(u) = \widehat F^{\star}_{\beta}(u)$ in \eqref{G-fun} yields \eqref{R.est.cens}; that is, ${\cal S}_{n,R}(\beta,\widehat F^{\star}_{\beta}(\cdot)) = \Psi_n(\beta).$ Substituting $\gamma(u) = \widehat F^\star_{\beta}(u)$ into \eqref{equiv-wt} and using \eqref{S-int} in the Appendix, it is also easily shown that
\begin{eqnarray*}
w(u;\beta) & = & \widehat F^\star_{\beta}(u) - \frac{1}{\widehat S_\beta(u)} \int_{u}^\infty \widehat F^\star_{\beta}(r )\, d \widehat F_{\beta}(r) 
= -\frac{1}{2} \widehat S_{\beta}(u-).
\end{eqnarray*}
Taken together, the results just summarized lead to the following set of equivalences:
$\Psi_n(\beta) = 
{\cal S}_{n,R}(\beta; \widehat F^{\star}_{\beta}(\cdot)) = 
{\cal S}_{n,T}(\beta; -\frac{1}{2}\widehat S^{\star}_{\beta}(\cdot-));$
that is,
\begin{eqnarray}
\label{WLR-Glen}
\Psi_n(\beta)  
\, \equiv \,
  -\frac{1}{2} \sum_{i=1}^n \int_{-\infty}^{\infty} \!\! \widehat S_{\beta}(u-)  ( X_i - \bar{X}(u; \beta) ) d N_i(u;\beta),
\end{eqnarray}
where the negative sign is required in this instance for mathematical equality when $\gamma(u) = \widehat F^\star_{\beta}(u),$ but is otherwise unimportant. In particular, the generalization \eqref{R.est.cens} of the R-estimator to right-censored data is a particular example of the class of estimating equations studied in both \citet{ritov1990} and \citet{tsiatis1990}. These results, in particular mathematical equivalence, rely heavily on the use of the mid-CDF $\widehat F^\star_{\beta}(\cdot).$

The selection $\gamma(u) = \widehat F^\star_{\beta}(u)$ does not satisfy all assumptions imposed on $\gamma(\cdot)$ in \citet{ritov1990}. For purposes of completeness, we therefore state the equivalence between \eqref{R.est.cens} and \eqref{WLR-Glen} below as a theorem, and provide an independent algebraic proof in Section \ref{app: Psi.eq.WLR} of the Appendix.
\begin{thm}
\label{thm:Psi.eq.WLR}
The estimating function
$\Psi_n(\beta)$ in \eqref{R.est.cens}
is equal to that given by \eqref{WLR-Glen}.
\end{thm}

\subsubsection{Asymptotics: a brief summary}
\label{sec:Wilcox-asy}
The equivalence between \eqref{R.est.cens} and \eqref{WLR-Glen} allows for relatively straightforward asymptotic developments to support inference for the proposed R-estimator, including several approaches to estimating the asymptotic variance of the estimated parameters. 
Suppose $\widehat \beta$ satisfies $n^{-1} \Psi_n(\widehat \beta) = o_P(n^{-1/2})$ \citep[e.g.,][]{Huang2013}, and that the regularity conditions given in \citet{Jin2004} hold; see also Appendix \ref{app: asycond}. These conditions include an assumption that implies $F$ is absolutely continuous with a differentiable  hazard function $\lambda_F(\cdot).$ To reduce the need for overly technical development, we additionally assume that $(T_i, \Delta_i,X_i),i = 1,\ldots,n$ are independent and identically distributed.  Define
\begin{equation}
\label{eq:sigma}
\Sigma = \frac{1}{4} \int_{-\infty}^{\infty} S^2(u-) {\mathcal H}(u) \Theta_0(u) d \Lambda_F(u),
\end{equation}
and
\begin{equation}
\label{eq:Xi}
\Xi = \frac{1}{2} \int_{-\infty}^{\infty} S(u-) {\mathcal H}(u) \Theta_0(u) \frac{\lambda_F'(u)}{\lambda_F(u)} d \Lambda_F(u) \: \:,
\end{equation} 
where $\Lambda_F(t) = \int_0^t \lambda_F(u) du$ is the cumulative hazard function for $F(\cdot),$
\[
{\mathcal H}(u) = \frac{\Theta_2(u)}{\Theta_0(u)} - \left(\frac{\Theta_1(u)}{\Theta_0(u)}\right)^{\otimes 2} 
\]
and
\[
\Theta_j(u) = \lim_{n \rightarrow \infty} \frac{1}{n} \sum_{i=1}^n E[  X_i^{\otimes j} Y_i(u,\beta_0) ] \:.
\]
Then, under the indicated conditions and an extra condition that is guaranteed to hold when  $w(u;\beta) = -\widehat S_{\beta}(u-)/2$ (e.g, see Appendix \ref{app: asycond}), we have $n^{-1/2} \Psi_n(\beta_0) \stackrel{d}{\rightarrow} N(0, \Sigma)$ as $n \rightarrow \infty;$
hence, $\Sigma$ is the asymptotic variance-covariance matrix of $n^{-\frac{1}{2}}\Psi_n(\beta_0)$.
Assuming $\Xi$ is positive definite, we further have that 
\[
\sqrt{n} ( \widehat \beta - \beta_0) \stackrel{d}{\rightarrow} N(0, \Xi^{-1} \Sigma \Xi^{-1}).
\]
Although the assumed continuity of $F(\cdot)$ implies that $S(u-)$ can be replaced by $S(u)$ and $d \Lambda_F(u)$ can be replaced by $\lambda_F(u) du,$  we will use these more general expressions for later convenience.

\subsubsection{A decomposition of $\Sigma$}
\label{sec:alt-var-est}

The matrix $\Sigma$ in \eqref{eq:sigma} has a familiar form, being derived directly from \eqref{WLR-Glen} using martingale arguments. However, it is less clear how $\Sigma$ relates to the asymptotic variance of \eqref{R.est.cens}, considered as an R-estimating equation, where one might expect the variance to depend more directly on the product of the variance of the ranks and the variance of covariates $X$. 
The result below provides an informative decomposition of $\Sigma$ that provides further insight into this connection.

\begin{thm}
\label{thm:new var}
Suppose the regularity conditions of the previous section hold, and define
$\mu_X$ and $\Sigma_X$ to be the marginal mean  and covariance matrix of $X$. 
Then, $\Sigma = \Sigma_1 - \Sigma_2,$ where the nonnegative definite matrices $\Sigma_i$ are respectively given by 
\[
\Sigma_1 = \frac{1}{4} \int_{-\infty}^{\infty} 
E\bigl[
I\{ E_{\beta_0} \geq u \} (X - \mu_{X})^{\otimes 2}
\bigr]
S^2(u) d F(u)
\]
and
\[
\Sigma_2 = \frac{1}{4} \int_{-\infty}^{\infty} 
(\mu_{X|E_{\beta_0}}(u) - \mu_{X})^{\otimes 2}
P\{ E_{\beta_0} \geq u \}
S^2(u) d F(u),
\]
where $\mu_{X|E_{\beta_0}}(u) = \Theta_1(u) / \Theta_0(u)
= E[ X | E_{\beta_0} \geq u].$ Moreover,
\begin{align}
\label{eq:sigma1}
\Sigma_1 & =  \frac{\Sigma_X }{12} 
-
\frac{1}{12} E\bigl[ (X - \mu_{X})^{\otimes 2}
S^3(E_{\beta_0}) \bigr].
\end{align}
\end{thm}

The derivation of the decomposition $\Sigma = \Sigma_1 - \Sigma_2$ as given by the first two displayed expressions
in the theorem is straightforward. The derivation of the alternative expression 
\eqref{eq:sigma1} for $\Sigma_1$ is less transparent and makes use of integration-by-parts. 

The reader is referred to Section \ref{app:new var proof} of the Appendix for further details and comment, where it is also shown that $\Sigma_1$ equals the asymptotic variance of $n^{-1/2} \Psi_n(\beta_0)$ provided that we replace $\widehat{\mathcal{R}}( \tilde E_{i,\beta_0} )$ with $R_{\beta_0,i}$. As shown
there, this interpretation of $\Sigma_1$ arises directly from the observation that $X_i$ and $R_{\beta_0,i}$ are uncorrelated for each $i.$ In combination, the presence of $\Sigma_2$ in the decomposition $\Sigma = \Sigma_1 - \Sigma_2$ therefore arises from the need to use $\widehat F_{\beta_0}(\cdot)$ in place of $F(\cdot)$ in defining $n^{-1/2} \Psi_n(\beta_0).$ Interestingly, because $\Sigma_2$ is generally positive definite, this 
substitution reduces asymptotic variance. 

Finally, in the absence of both censoring and tied observations, 
observe that $\Sigma_1$ reduces to $\Sigma_X / 12,$ which is the asymptotic variance of \eqref{R.est} at $\beta = \beta_0$ when $a(u) = u$ \citep[e.g.,][Problem 5.5.3]{Hettmansperger84}. 

\subsection{R-Estimation and Inference in Practice}
\label{sec:variance}

Inference for $\beta_0$ is generally conducted on the basis of the asymptotic results summarized in Section \ref{sec:Wilcox-asy}. In particular, Wald-style confidence intervals and tests for $\beta_0$ can be constructed in the usual way given suitable estimates of both $\beta_0$ and $\Omega$; conversely, the quasi-score test for $\beta_0$ only requires an appropriate estimate of $\Sigma.$  

Outside of the quasi-score test, there are some well-known challenges in using such results for both estimation and inference, primarily stemming from (i) the need to find a solution $\widehat \beta$ that lies within a $n^{-1/2}-$neighborhood of $\beta_0$ due to the lack of monotonicity in $\Psi_n(\beta);$ and, (ii) the presence of $\lambda'_F/\lambda_F$ in \eqref{eq:Xi}, which is generally difficult to estimate well. We discuss these two issues in the following sections, after first considering estimation of $\Sigma$ and the quasi-score test.

\subsubsection{Estimating $\Sigma$ and the Quasi-Score Test}

The quasi-score test, based on $\Psi_n(\beta)$ as given in \eqref{WLR-Glen} (equivalently, \eqref{R.est.cens}), is equivalent to a censored data linear rank test that uses the Wilcoxon score and may be preferred over the Wald test due to its fewer estimation requirements. The quasi-score test statistic for $H_0: \beta_0 = 0$ satisfies $n^{-1} [\Psi_n(0)]^T \Sigma^{-1} \Psi_n(0) \rightarrow \chi^2_p$ under $H_0$ and, advantageously, is also locally most powerful in the case where $\epsilon$ has a logistic distribution; see \cite{prentice1978} for further discussion. In this test, $\Sigma$ can be replaced by $\widehat \Sigma(0),$ defined via the following 
plug-in estimator for $\Sigma:$
\begin{equation}
\label{eq:WLR-sig}
\widehat \Sigma(\beta) = \frac{1}{4} \int_{-\infty}^{\infty} \widehat S^2_{\beta}(u-) \widehat {\mathcal H}(u;\beta) \widehat \Theta_0(u;\beta) d \widehat \Lambda_{\beta} (u)
\end{equation}
where
\[
\widehat {\mathcal H}(u;\beta) = \frac{\widehat \Theta_2(u;\beta)}{\widehat \Theta_0(u;\beta)} - \left(\frac{\widehat \Theta_1(u;\beta)}{\widehat \Theta_0(u;\beta)}\right)^{\otimes 2}, 
\]
\[
\widehat \Theta_j(u;\beta) = \frac{1}{n} \sum_{i=1}^n  X_i^{\otimes j} Y_i(u,\beta),
\]
and
\begin{equation}
\label{eq:Lam-hat}
\widehat \Lambda_{\beta} (u)
= \frac{1}{n} \sum_{i=1}^n \int_{-\infty}^u \frac{ d N_i(s,\beta)}{\widehat \Theta_0(s;\beta)}.
\end{equation}
The formula \eqref{eq:WLR-sig} simplifies considerably: 
\begin{equation}
\label{sig-hat-WLR}
\widehat \Sigma(\beta) = \frac{1}{4n} \sum_{i=1} \Delta_i [ \widehat S_{\beta}(\tilde E_{i,\beta}-)
]^2 \widehat {\mathcal H}(\tilde E_{i,\beta};\beta).
\end{equation}
In place of $\widehat \Sigma(0),$ one can instead use $\widehat \Sigma = \widehat \Sigma(\widehat \beta)$ for testing the global null hypothesis that $\beta_0 = 0$ at the expense of increased computational burden. More generally, score-type tests for composite hypotheses can be constructed using hypothesis-restricted estimators of $\beta_0$.

\subsubsection{Estimating $\beta_0$}

Estimating $\beta_0$ really involves two related problems. One problem involves locating a solution in the indicated $n^{-1/2}-$neighborhood of $\beta_0,$ and this is often addressed in the literature by making use of the well-known fact that the Gehan-weighted logrank estimating function corresponds to the (sub)gradient of a convex objective function and is (under certain regularity conditions) guaranteed to produce a $\sqrt{n}-$consistent initial estimator $\widehat \beta_G.$ Once found, $\widehat \beta_G$ can be used as a starting point for determining a $\widehat \beta$ such that $n^{-1}  \Psi_n(\widehat \beta) = o_P(n^{-1/2}).$  The other problem involves actually finding such a $\widehat \beta,$ that is, a solution where \eqref{R.est.cens} ``equals'' zero. This root-finding problem is challenging mainly because \eqref{R.est.cens} lack smoothness as a function of $\beta$. Indeed, even given $\widehat \beta_G,$ standard root-finding packages may not perform well because the resulting solution is typically only a ``zero crossing" of \eqref{R.est.cens}, rather than an exact zero.

For the simulation work done later in this paper, we devised and used a new sequential univariate equation solver that makes repeated use of a simple bisection algorithm; this algorithm easily identifies zero crossings in any one-dimensional root-finding problem. Over the course of numerous simulation studies, no difficulties were encountered when estimating parameters using $\beta=0$ as a starting value. This is not to say that making use of $\widehat \beta_G$ is not necessary and/or advisable in general, only that the problems we considered ultimately did not require us to increase the overall computational burden by first determining $\widehat \beta_G$. In cases where $\widehat \beta_G$ is desired or otherwise useful, a number of {\sf R} packages now exist that compute or otherwise approximate $\widehat \beta_G$ should it be needed. Related ``induced smoothing'' methods can also significantly ease the computational burden of determining $\widehat \beta_G$ in larger problems; see, for example, \cite{Brown2005} and \citet{johnstraw09}.  \citet{Huang2013} provides yet another approach. 

\subsubsection{Estimating $\Omega$, Wald Tests and Confidence Intervals}
\label{sec:MC-var-est}

The estimation of $\Omega = \Xi^{-1} \Sigma \Xi^{-1},$ the asymptotic covariance matrix of $\sqrt{n} ( \widehat \beta - \beta_0),$
is more challenging than estimating $\Sigma$ because the ``slope'' matrix $\Xi$ cannot be easily obtained as the first derivative of 
the non-smooth estimating function $\Psi_n(\beta).$  A comparatively recent Monte Carlo least-squares procedure for estimating $\Omega$ that avoids this problem is proposed in \citet{zeng-lin-2008}; see also \citet{Strawderman2005}, who developed an earlier variant for the accelerated failure time model applied to recurrent events. An important feature of these two approaches: only $\widehat \Sigma = \widehat \Sigma(\widehat \beta)$ is needed, and once calculated, an estimator $\widehat \Xi$ can then be obtained by exploiting the asymptotic linearity of $\Psi_n(\beta)$ through a combination of Monte Carlo sampling and linear regression. In particular, one can estimate $\Xi,$ and thus $\Omega,$ as follows:
\begin{enumerate}
\item Generate $\tilde \beta_b = \widehat \beta + Z_b / \sqrt{n}$ where $Z_b$ is a $p-$vector
with mean zero and a user-specified $p \times p$ covariance matrix ${\cal D}_Z,$  $b = 1,\ldots,B$.
\item Compute $n^{-1/2} \Psi_n(\tilde \beta_b), b = 1,\ldots, B$. 
\item Center $n^{-1/2} \Psi_n(\tilde \beta_b)$ and $\tilde \beta_b - \widehat \beta$ by their respective
sample means, compute $\tilde \Xi$ by regressing the former on the latter, and estimate $\Xi$ by
$\widehat \Xi = 0.5 (\tilde \Xi + \tilde \Xi^T).$
\item Estimate $\Omega,$ the covariance matrix of $\sqrt{n}(\widehat \beta - \beta_0),$ by $\widehat \Xi^{-1} \widehat \Sigma
\widehat \Xi^{-1}.$
\end{enumerate}
The primary advantage of this methodology is that it avoids the need to repeatedly find zeros 
of a nonsmooth, possibly non-monotone estimating equation. The primary disadvantage is that it 
can inject some instability by introducing an additional source of variability (i.e., 
Monte Carlo approximation error) that depends in part on the user-specified
variance matrix ${\cal D}_Z.$ \citet{zeng-lin-2008} are not specific as to how to best choose 
${\cal D}_Z.$ In practice, it is important to select ${\cal D}_Z$ so that $\widetilde\beta_b, b = 1,\ldots,B$ 
generally fall within the region about $\widehat\beta$ where $\Psi_n(\beta)$ is approximately linear. 
Although this is guaranteed to occur asymptotically, the estimator $\widehat \Xi$ obtained in Step 3 
may not always perform well, especially for smaller sample sizes, or when ${\cal D}_Z$ is chosen poorly. 

An alternative procedure for estimating $\Omega$ 
can be derived using the ``inverse numerical differentiation'' approach proposed in \citet{Huang2002}.  
Like the approach(es) described above, Huang's method only requires the availability of $\widehat \Sigma,$
and exploits asymptotic linearity for justification. However, it does not rely on Monte Carlo sampling, 
instead requiring $p$ additional root-finding steps. Below, we propose a minor variant of Huang's 
original approach. Let the symmetric square root of $n^{-1} \widehat \Sigma$ be given by $\widehat{\mathcal{C}}$.  
For each column $\widehat{\mathcal{C}}_{k}$ of $\widehat{\mathcal{C}},$ we proceed by first finding a 
`positive' solution $\widehat\beta^{(k)}_u$ that solves
\begin{equation}\label{pos.soln}
    n^{-1}\Psi_n(\beta^{(k)}_u)= \widehat{\mathcal{C}}_{k}
\end{equation}
and then determine a corresponding `negative' solution $\widehat\beta^{(k)}_l$ that solves
\begin{equation}\label{neg.soln}
    n^{-1} \Psi_n(\beta^{(k)}_l)=-\widehat{\mathcal{C}}_{k}.
\end{equation}
Since $\widehat{\mathcal{C}}_{k}$ is $O_p(n^{-\frac{1}{2}})$ while $n^{-1} \Psi_n(\widehat \beta)$ is $o_p(n^{-1/2})$, each of the solutions $\widehat \beta^{(k)}_u$ and $\widehat \beta^{(k)}_l$ should be close to $\widehat\beta,$ and can be numerically determined
in a comparatively efficient way. Next, we form the matrix $\widehat{\mathcal{B}},$ which is defined to have $k$th column given  $(\widehat{\mathbf{\beta}}^{(k)}_u- \widehat{\mathbf{\beta}}^{(k)}_l)/2$ for $k=1,\ldots, K.$ 
Results in Section 3.2 of \cite{Huang2002} imply that $\widehat{\mathcal{B}}$ consistently estimates 
$n^{-1/2} \mathbf{\Xi}^{-1} \Sigma^{1/2};$ consequently, $\Omega$
can be estimated by
\begin{equation}\label{eq:Omega.hat.huang}
    \widehat{\Omega}=n\widehat{\mathcal{B}}
    \widehat{\mathcal{B}}^T.
\end{equation}

In the event that no solution to \eqref{pos.soln} exists but \eqref{neg.soln} can be solved, we use $\widehat\beta-\widehat\beta^{(k)}_l$ for the $k$th column of $\widehat{\mathcal{\boldsymbol{B}}};$ similarly, if \eqref{pos.soln} can be solved but not \eqref{neg.soln}, we use $\widehat\beta^{(k)}_u-\widehat\beta$.

As will be seen later, the estimator $\widehat\Omega$ in \eqref{eq:Omega.hat.huang} performed well in our simulations. In part, this is because $\widehat{\mathcal{C}}_{k}$ defines an appropriate scale of deviation for $n^{-1}\Psi_n(\beta);$ solving $n^{-1}\Psi_n(\beta^{(k)})= \pm \widehat{\mathcal{C}}_{k}$ for $\beta^{(k)}$ then implicitly determines an appropriately scaled deviation $\beta^{(k)} - \widehat \beta.$ Importantly, though, the calculation of $\widehat\Omega$ (and also $\widehat \beta$) does become increasingly cumbersome as the number of parameters increases (i.e., due to the dependence of this method on repeatedly applying a root-finding procedure). For data with many parameters, the Monte-Carlo approach summarized above may prove to be more computationally attractive, provided one can estimate $\beta_0$ and select an appropriate choice of ${\cal D}_Z.$ The variant of this Monte Carlo method proposed in \citet{Strawderman2005}, mentioned earlier, provides one possible solution to this problem by sampling in a neighborhood of $\widehat \beta_G$ using its estimated covariance matrix instead of a neighborhood of $\widehat \beta$ defined by  the arbitrary matrix ${\cal D}_Z.$

Given a suitable estimator $\widehat\Omega$, Wald tests having asymptotic normal or $\chi^2$ distributions can be constructed in the usual way.  For example, $H_0:\beta=\beta_0$ can be tested using $n(\widehat\beta-\beta_0)^T\widehat\Omega^{-1}(\widehat\beta-\beta_0)$ which has an asymptotic $\chi^2_p$ distribution.

\section{Extending the R-estimator to general scores}
\label{sec:nonlinear score}
The results of the previous sections focus on the linear function $a(u) = u$ where $u \in [0,1].$ Many authors have previously considered the problem of estimating $\beta$ using \eqref{R.est} for a nonlinear choice of $a(\cdot);$ see, for example, \citet{Hettmansperger84,Hettmansperger2e} in the case of uncensored data. In the context of linear rank tests, \citet{prentice1978} considers several possible choices of $a(\cdot),$ including the Wilcoxon and so-called ``logrank'' scores derived from the extreme value distribution; more generally, one can define  $a(u) = -f'_0(F^{-1}_0(u))/f_0(F^{-1}_0(u))$  for some hypothesized PDF $f_0(\cdot)$ (i.e., for $\epsilon$) and its associated CDF $F_0(\cdot)$ \citep[e.g.,][]{Lai-Ying-92}, or equivalently parameterized through $S_0(\cdot) = 1- F_0(\cdot)$ \citep[e.g.,][]{prentice1978, cuzick85}. This particular choice of score function is asymptotically optimal when $F_0$ and $f_0$ correspond to the true distribution of $\epsilon.$ Below, we show how to extend the results of  the previous sections to the case of a right-censored outcome. We first consider the case of bounded score functions; then, we discuss the case where the score functions may be unbounded.

\subsection{Bounded score functions}

Let $a: [0,1] \rightarrow \mathbb{R}$ be a given continuously differentiable monotone increasing (possibly nonlinear) score function, and define $A(u) = \int_0^u a(s) ds$ for $u \in (0,1].$ Note that $a(u) = u$ implies $A(u) = u^2/2,$ which corresponds to the mid-rank $n  \widehat{\mathcal{R}}( \tilde E_{i,\beta} ) $ as defined through \eqref{est.rank.3}. More generally, if one defines $a(u)$ using 
\begin{equation}\label{eq:choice.for.a}
    a(u) = -\frac{f'_0(F^{-1}_0(u))}{f_0(F^{-1}_0(u))}
\end{equation}
for a user-specified choice of $f_0(\cdot),$  it is easy to show that 
\begin{equation}\label{eq:A.closed.form}
    A(u)=f_0(F^{-1}_0(0))-f_0(F^{-1}_0(u))
\end{equation}
so that $A(u)$ is easily computed provided that $F^{-1}_0(\cdot)$ readily available.
In this section, $a(u)$ is assumed bounded for $u \in [0,1];$ discussion of the unbounded case is deferred to the next subsection.

The simple form of \eqref{est.rank.3} relies directly on the use of the mid-rank distribution $\widehat F^{\star}(\cdot)$ in both \eqref{est.rank.bj} and  \eqref{est.rank.2}, the latter being especially critical to both the simplification of the integral term appearing in  \eqref{est.rank.bj}  and the rank-sum preservation result in Theorem \ref{nice result 2}. It is therefore natural to ask whether the mid-CDF $\widehat{F}^{\star}( \cdot )$ plays a similarly central role in generalizing \eqref{est.rank.bj} to the case of a nonlinear function $a(\cdot);$ that is, whether 
\[
\Delta_i  a(\widehat{F}^{\star}_{\beta}( \tilde E_{i,\beta} )) + (1-\Delta_i)
\int_{ \tilde{E}_{i,\beta}}^\infty a(\widehat{F}^{\star}_{\beta}(y)) \frac{d\widehat{F}_{\beta}(y)}{\widehat{S}_{\beta}(\widetilde{E}_{i,\beta})},
\]
is indeed the most appropriate generalization of \eqref{est.rank.bj}. The answer here is negative, at least in the sense of preserving exact (versus asymptotic) correspondence; although well-defined, this expression fails to similarly generalize \eqref{est.rank.bj} because
\[
\int_{ \tilde{E}_{i,\beta}}^\infty a(\widehat{F}^{\star}_{\beta}(y)) \frac{d\widehat{F}_{\beta}(y)}{\widehat{S}_{\beta}(\widetilde{E}_{i,\beta})}
\neq 
\frac{A(1)-A(\widehat{F}_{\beta}(\widetilde{E}_{i,\beta}))}{\widehat{S}_{\beta}(\widetilde{E}_{i,\beta})}.
\]
This integral equality fails in general because $\widehat{F}_{\beta}(\cdot)$ only has a generalized inverse. 

One might therefore ask whether there exists any function, say $\widehat F^{\star}_{a,\beta}(t),$ such that 
\begin{align}
\label{eq:gen-Wilcoxon-2}
\widehat{\mathcal{R}}^{(a)} & ( \tilde E_{i,\beta} )  = \Delta_i a\bigl(\widehat F^{\star}_{a,\beta}(\tilde E_{i,\beta})\bigr)  + (1-\Delta_i) 
\frac{1}{\widehat S_{\beta}(\tilde E_{i,\beta})} \int_{\tilde E_{i,\beta}}^{\infty} a\bigl(\widehat F^{\star}_{a,\beta}(s)\bigr) \, d \widehat F_{\beta}(s) \\
\label{eq:gen-Wilcoxon-1}
& = 
\Delta_i a\bigl(\widehat F^{\star}_{a,\beta}(\tilde E_{i,\beta})\bigr)  + (1-\Delta_i) 
\left[ 
\frac{A(1) - A( \widehat F_{\beta}(\tilde E_{i,\beta} ) )}{\widehat S_{\beta}(\tilde E_{i,\beta} )} \right];
\end{align}
note that equality here requires
\begin{equation}
\label{eq:key-integral}
\int_{t}^{\infty} a\bigl(\widehat F^{\star}_{a,\beta}(s)\bigr) \, d \widehat F_{\beta}(s) = A(1) - A(\widehat F_{\beta}(t)).
\end{equation}
Were $\widehat F_{\beta}(t)$ replaced by a continuous function of $t,$ the equality in \eqref{eq:key-integral} can easily be shown to hold 
\citep[e.g.,][]{falkner12}. More generally, we will show below that an affirmative answer also exists for the step function $\widehat F_{\beta}(\cdot).$ In particular, for each $t,$ let  $\widehat F^{\star}_{a,\beta}(t)$ be defined through the equality
\begin{equation}
\label{eq:Fhat-a}
    a(\widehat{F}^{\star}_{a,\beta}(t))=\frac{A( \widehat  F_{\beta}(t))  - A( \widehat  F_{\beta}(t-) )}{\widehat  F_{\beta}(t)  - \widehat  F_{\beta}(t-)};
\end{equation}
a unique solution $\widehat{F}^{\star}_{a,\beta}(t)$ is guaranteed to exist by the Mean Value Theorem
because $a(\cdot)$ is continuous and monotone. It is now easy to see that \eqref{eq:key-integral} simplifies:
\begin{equation*}
\int_{t}^{\infty} a\bigl(\widehat F^{\star}_{a,\beta}(s)\bigr) \, d \widehat F_{\beta}(s) = 
\sum_{s > t} 
a\bigl(\widehat F^{\star}_{a,\beta}(s)\bigr) \times \bigl( \widehat  F_{\beta}(s)  - \widehat  F_{\beta}(s-) \bigr) =
\sum_{s > t} \bigl( A(\widehat  F_{\beta}(s))  - A(\widehat  F_{\beta}(s-))\bigr),
\end{equation*}
the last summation telescoping to $A(1) - A(\widehat F_{\beta}(t))$ because the self-consistent estimator satisfies $\widehat  F_{\beta}(\infty) = 1.$ Consequently, the preservation of \eqref{eq:key-integral} requires using a specific generalization of $\widehat{F}^{\star}_{\beta}(t)$ that depends on the choice of $a(\cdot).$  Fortunately, use of \eqref{eq:gen-Wilcoxon-1} in practice does not require the explicit calculation of $\widehat{F}^{\star}_{a,\beta}(t)$, as the relationship given in \eqref{eq:Fhat-a} can be used directly.

The results developed above show that one can now estimate $\beta$ by determining a solution to 
$\Psi_n(\beta;a) = 0,$ where
\begin{equation}\label{R.est.cens.gen}
    \Psi_n(\beta;a) = \sum_{i=1}^n  \widehat{\mathcal{R}}^{(a)} ( \tilde E_{i,\beta} ) ( X_{i} - \bar X)
\end{equation}
and \eqref{eq:gen-Wilcoxon-1} and \eqref{eq:Fhat-a} combine to give
\begin{align*}
\widehat{\mathcal{R}}^{(a)}(\tilde E_{i,\beta} ) & = \Delta_i 
\biggl( \frac{A( \widehat  F_{\beta}(\tilde E_{i,\beta} ))  - A( \widehat  F_{\beta}(\tilde E_{i,\beta} -) )}{\widehat  F_{\beta}(\tilde E_{i,\beta} )  - \widehat  F_{\beta}(\tilde E_{i,\beta} -)}
\biggr) + (1-\Delta_i) 
\left[ 
\frac{A(1) - A( \widehat F_{\beta}(\tilde E_{i,\beta} ) )}{\widehat S_{\beta}(\tilde E_{i,\beta} )} \right].
\end{align*}
It is easily shown that $\widehat{\mathcal{R}}^{(a)}(\tilde E_{i,\beta} )$ reduces to \eqref{est.rank.bj} when $a(u) = u$
and $A(u) = u^2/2,$ and hence that  \eqref{R.est.cens.gen} reduces to \eqref{R.est.cens}.

The above development relies on two features of the self-consistent estimator $\widehat F_{\beta}(\cdot)$: (i) it is a step-function; and, (ii) for each finite $\beta$, it leads to a proper discrete CDF.  We now show that the key elements of these arguments generalize to any arbitrary proper CDF $H(\cdot).$ Such developments, interesting in their own right, also lay useful ground work for generalizing both 
Theorems \ref{nice result} and \ref{nice result 2}.
Let
$U \sim \mbox{Unif}(0,1)$ and define for any $t$ 
\begin{equation*}
\gamma_{a}(t;H)  = E\left[ a( U H(t) + (1-U) H(t-) ) \right]
\end{equation*}
where the expectation is taken over $U$. The function $\gamma_a(t;H)$ may not seem very intuitive at first. However, the following observations are immediate:
\begin{itemize}
\item When $H(\cdot)$ is continuous and $a(\cdot)$ is non-linear, we have $\gamma_{a}(t;H) = a(H(t));$

\item For general $H(\cdot)$ and $a(u) = u,$ we have $\gamma_{a}(t;H) = H^{\star}(t),$ where $H^{\star}(\cdot)$ is the corresponding mid-CDF; if $H(\cdot)$ is also continuous, $\gamma_{a}(t;H) = H(t);$
\end{itemize}
Hence, $\gamma_a(t;H)$ reduces to $a(H^{\star}(t))$ in several important use cases, though is distinct from $a(H^{\star}(t))$ more generally. Importantly, $\gamma_a(t;H)$ also has a simple closed form representation: for each fixed $H(t)$ and $t,$
\begin{eqnarray}
\label{eq:gamma_a}
\gamma_{a}(t;H) & = & 
\int_0^1 \!\! a( s H(t) + (1-s) H(t-) ) ds 
\, = \, 
\frac{A( H(t))  - A( H(t-) )}{H(t)  - H(t-)},
\end{eqnarray}
where the right-hand side is to be interpreted as being equal to $a(H(t))$ when $H(\cdot)$ is continuous at $t$.
For the particular choice $H(t) = \widehat F_{\beta}(t),$ 
we have
$\gamma_{a}(t;\widehat F_{\beta}) = a(\widehat{F}^{\star}_{a,\beta}(t))$ as given in
\eqref{eq:Fhat-a}, which also reduces to the mid-CDF $\widehat{F}^{\star}_{\beta}(t)$ when
$a(u) = u.$  Thus, \eqref{eq:gamma_a} generalizes \eqref{eq:Fhat-a} to an arbitrary proper CDF $H$.  Critically, if one then defines
\begin{equation}
\label{eq:Gamma_a_1}
\Gamma_{a}(t;H) = \frac{1}{1-H(t)} \int_{t}^{\infty} \gamma_{a}(s;H) \, d H(s),
\end{equation}
results in \citet[Thm.\ 7.2]{genchainrule} that generalize the chain and substitution rules for Stieltjes integrals 
further imply $\int_{t}^{\infty} \gamma_a(s;H) d H(s)  = A(1) - A(H(t)),$  generalizing \eqref{eq:key-integral}. Consequently, for $t$ such that $H(t) < 1,$
\begin{equation}
\label{eq:Gamma_a_2}
\Gamma_{a}(t;H) = \frac{A(1 ) - A(H(t))}{1-H(t)};
\end{equation}
because $a(\cdot)$ is bounded and continuously differentiable, it can be shown that \eqref{eq:Gamma_a_2}
converges to $a(1)$ as $H(t) \rightarrow 1.$

Theorem \ref{nice result 2} establishes a nice result on the sum of \eqref{est.rank.3}, equivalently \eqref{est.rank.bj}. The following result uses \eqref{eq:gamma_a} and the equivalence of \eqref{eq:Gamma_a_1} and \eqref{eq:Gamma_a_2} to 
generalize both Theorems  \ref{nice result} and  \ref{nice result 2} to the case of a nonlinear $a(\cdot).$ 
\begin{thm}
\label{really nice result}
Define $F$ to be the CDF of $\epsilon,$ and let
$
\mathcal{R}^{(a)}_{\beta_0} = \Delta \gamma_a(\tilde E_{\beta_0};F) + (1-\Delta) \Gamma_a(\tilde E_{\beta_0};F)
$
where 
$\tilde E_{\beta_0} = Y - X^T \beta_0$ and $\Delta$ are respectively the true observed residual and failure status
and $F$ is the CDF of $\epsilon = \log T^* - X^T \beta_0.$
Then,  $E(\mathcal{R}^{(a)}_{\beta_0} \mid X) = E(\mathcal{R}^{(a)}_{\beta_0}) = A(1) - A(0).$
Moreover, when the self-consistent estimator of $F$ is used and $\| \beta \| < \infty,$
\begin{equation}
\label{est.ranksum}
\sum_{i=1}^n \widehat{\mathcal{R}}^{(a)} ( \tilde E_{i,\beta} )
= n ( A(1) - A(0) ),
\end{equation}
where $\widehat{\mathcal{R}}^{(a)} ( \tilde E_{i,\beta} )$ is defined as in 
\eqref{eq:gen-Wilcoxon-1}.
\end{thm}
Theorem \ref{really nice result} is proved in Appendix \ref{app: thm5proof}. To see that Theorem \ref{really nice result} contains Theorems \ref{nice result} and \ref{nice result 2} as special cases, recall that $a(u) = u$ implies  $A(u) = u^2/2;$ hence, $A(1)-A(0) = 1/2.$

Following earlier developments, we conjecture that \eqref{R.est.cens.gen} has an exact counterpart within Tsiatis' class of weighted logrank statistics. Recall that the equivalence between the Ritov and Tsiatis classes of estimating equations relies on selecting a weight of the form \eqref{equiv-wt}. Selecting $\gamma(\cdot) = \gamma_{a}(t;\widehat F_{\beta})$ in \eqref{equiv-wt} and using the equivalence between
\eqref{eq:Gamma_a_1} and \eqref{eq:Gamma_a_2}, the estimating equation \eqref{R.est.cens.gen}  should equal \eqref{T-fun} with 
weight function $w(u;\beta) = w_{\beta,a}(u),$ where
\begin{equation}
\label{eq:exact-weight}
w_{\beta,a}(u) = \frac{A( \widehat F_{\beta}(u))  - A( \widehat F_{\beta}(u-) )}{\widehat F_{\beta}(u)  - \widehat F_{\beta}(u-)}
- \frac{A(1) - A( \widehat F_{\beta}(u ) )}{\widehat S_{\beta}(u )};
\end{equation}
as noted just below \eqref{eq:Gamma_a_2}, the second term on the right-hand side of \eqref{eq:exact-weight} reduces to $a(1)$ 
for all $u$ such that $\widehat F_{\beta}(u ) = 1.$
Under mild additional conditions on $a(\cdot)$, the large sample theory and methods for inference and variance estimation respectively 
described in Sections \ref{sec:Wilcox-asy} and \ref{sec:variance} can be extended  to the case of more general scores.
For example, in the variance estimation procedure of Section \ref{sec:MC-var-est}, one can use an estimator of the 
variance of \eqref{R.est.cens.gen} that generalizes \eqref{sig-hat-WLR} to a nonlinear $a(\cdot);$
that is, $\widehat \Sigma = \widehat \Sigma(\widehat \beta)$ can be replaced by \textcolor{blue}{$\widehat\Sigma_a(\widehat\beta)$ where}
\begin{equation}
\label{sig-hat-WLR.gen}
\widehat \Sigma_a(\beta) = \frac{1}{n} \sum_{i=1} \Delta_i [ w_{\beta,a}( \tilde E_{i,\beta} ) ]^2 
\widehat {\mathcal H}(\tilde E_{i,\beta};\beta).
\end{equation}
It is further conjectured that the use of any consistent estimator of $F$  will not  change the asymptotic behavior of $\sqrt{n}(\widehat \beta - \beta)$ or its variance estimator provided conditions analogous to those used in Section \ref{sec:Wilcox-asy} hold. For example, using
\begin{equation*}
\widetilde w_{\beta,a}(u) = a(\widehat F^{\star}_{\beta}(u)) -
\frac{A(1) - A( \widehat F_{\beta}(u ) )}{\widehat S_{\beta}(u )}
\end{equation*}
in place of \eqref{eq:exact-weight} should not change asymptotic behavior, at  least when the underlying distribution $F$ is continuous.
Further, we expect that the results of Section \ref{sec:alt-var-est} can be extended such that the analog of $\Sigma_1$ represents the 
the asymptotic variance of $n^{-1/2} \Psi_n(\beta_0;a)$ when $\widehat F_{\beta}(u )$ is replaced
by $F(u)$ in \eqref{R.est.cens.gen}.

Given the variance estimator \eqref{sig-hat-WLR.gen}, the pseudo-score statistic for general weight function $a(u)$, given by $n^{-1}[\Psi_n(0,a)]^T\widehat\Sigma^{-1}_a(0)\Psi_n(0,a),$ will have an asymptotic $\chi^2$ distribution under $H_0:\beta_0=0$, and is the locally most powerful test in the case where $a(u)$ is the optimal score corresponding to the distribution of $\epsilon$.  Further, either of the methods described in \ref{sec:MC-var-est} can be used to obtain $\widehat\Omega_a$, the estimated variance-covariance matrix of $\sqrt{n}(\widehat\beta-\beta)$ that applies when weight function $a(\cdot)$ has been used, and can be used to construct
Wald confidence intervals and Wald tests in the usual way. In our experience to date, Huang's approach performed as well or better that the indicated  Monte Carlo procedure for estimating $\Omega_a.$  

\begin{remark}
Define
\[
w_a(u) = a(u) - \frac{A(1) - A(u) }{S(u)};
\]
then, \citet[Sec.\ 3]{cuzick85}  shows that \eqref{T-fun},  with $w(u;\beta) = w_a(\widehat F_{\beta}(u)),$ results in a rank test statistic that is asymptotically equivalent to the class of linear rank statistics originally proposed in \citet{prentice1978}. This particular choice of weight, and corresponding equivalence result, essentially anticipates a key observation made in \citet{ritov1990}. In particular, selecting $\gamma(u) = a(\hat F_{\beta}(t))$ in Ritov's estimating function leads to a weighted logrank statistic \eqref{T-fun} with $w(u;\beta) = W_F a(\widehat F_{\beta}(u)),$ where
\begin{eqnarray*}
W_F a(\widehat F_{\beta}(u)) & = & a(\widehat F_{\beta}(u)) - \frac{\int_u^{\infty} a(\widehat F_{\beta}(s)) d \widehat F_{\beta}(s)}{1-\widehat F_{\beta}(u)} \\
& \approx & 
a(\widehat F_{\beta}(u)) -
\frac{\int_{\widehat F_{\beta}(t)}^{1} a(s) d s}{1-\widehat F_{\beta}(u)} \\
& = & 
a(\widehat F_{\beta}(u)) -
\frac{A(1) - A( \widehat F_{\beta}(u) ) }{\widehat S_{\beta}(u)}
\\
& = & w_a(\widehat F_{\beta}(u)).
\end{eqnarray*}
The fact that the step-function $\widehat F_{\beta}(u)$ only has a 
generalized inverse is the reason that the transition between the first and second lines involves an approximation.
\end{remark}

\begin{remark}
The use of the self-consistent estimator implies that $\widehat F_{\beta}(u)$ and $\widehat F_{\beta}(u-)$ both equal 1 for values of $u$ that exceed the largest observed residual. Under the condition that $a'(u)$ remains bounded for $u \in (0,1],$ one can use integration-by-parts to prove that $\gamma_{a}(u;\widehat F_{\beta}) \rightarrow a(1)$ and $\Gamma_{a}(u;\widehat F_{\beta}) \rightarrow a(1),$ and hence that $w_{\beta}(u) \rightarrow 0,$ as $\widehat F_{\beta}(u) \rightarrow 1$. 
\end{remark}

\begin{remark}
Under the assumption that $a(\cdot)$ is twice continuously differentiable with a bounded third derivative, 
one can do a Taylor series expansion of $A(H(t))$ at $H(t) = H^{\star}(t);$ similarly, one can do a Taylor series expansion of
$A(H(t-))$ at $H(t-) = H^{\star}(t).$ Doing so and subtracting the two expansions, it can be shown that
\[
\gamma_a(t;H) = a(H^{\star}(t)) + \frac{1}{24} a''(H^{\star}(t)) (H(t)-H(t-))^2 + O( (H(t)-H(t-))^4 ).
\]
Hence, $\gamma_a(t;H)$ can be well-approximated by $a(H^{\star}(t))$ up to an error term that
depends on both the magnitude of $(H(t) - H(t-))^2$ and the sign and magnitude of $a''(H^{\star}(t)).$ 
\end{remark}

\subsection{Unbounded score functions}

The developments of the previous section assume that $a(\cdot)$ is bounded in addition to satisfying some mild continuity and differentiability conditions. In the context of linear rank tests, two common examples of unbounded score functions include $a(u) = \Phi^{-1}(u)$ (i.e., normal scores) and $a(u) = -1-\log(1-u)$ (i.e., extreme minimum value, or logrank, scores).  If it is desirable to consider the use of an unbounded score function, one practical solution that will bound evaluation of $a(u)$ away from 
$u=0$ and $u=1$ is to replace it by $a([nu+0.5]/[n+1]),$ for normal scores or $a([nu]/[n+1])$ for extreme value scores, where $n$ is the sample size.
Asymptotic results have previously been established for the class of linear rank tests (and extensions
to regression) with unbounded scores under somewhat more restrictive conditions on the tails of the  error distribution (i.e., compared to the case of bounded score); see, for example, \citet{JJ1972} in the case of complete data and \citet{Lai-Ying-92}  in the case of right-censored (possibly left-truncated) data. However, as noted in \citet[Sec.\ 3.6.3]{Hettmansperger2e},
the use of an unbounded score function $a(\cdot)$ will lead to an R-estimator for $\beta$ with 
uncensored data that has an unbounded influence function, reducing robustness in the presence of more
extreme residuals.  The use of bounded scores is therefore recommended to enhance robustness.

Another simple approach to deriving a bounded score from an unbounded score is to ``Winsorize'' the latter. For example,
defining $z_{1-\alpha}$ so that $\Phi(z_{1-\alpha}) = 1-\alpha$ for $\alpha \in [0,1/2],$ the normal score $a(u) = \Phi^{-1}(u)$ can be Winsorized by respectively setting it equal to $-z_{1-\alpha}$ for $u < \alpha$ or $z_{1-\alpha}$ for $u > 1-\alpha.$  Alternatively, and more generally,
\citet{McKean-Sievers-89}  proposed to use the generalized $F-$distribution \citep[e.g.,][Sec.\ 2.2.7]{KalbfleischPrentice2e} with finite choices of the degrees-of-freedom parameters to generate bounded scores suitable for regression parameter
estimation with uncensored data in cases where $F$ may be asymmetric. Under its original parameterization, the generalized $F_{m_1,m_2}$ distribution refers to the distribution of the failure time $T^*$, and corresponds to a location-scale model for
\eqref{linear.model} in which the residual $\epsilon$ follows natural log of a $F_{2m_1,2 m_2}$ random variable (i.e., up
to a scale parameter $\sigma > 0$). The generalized $F-$distribution is highly flexible. On the scale of $\epsilon$, it contains the logistic distribution as a special case; moreover, it generates numerous other distributions, including the extreme value distributions for minima or maxima, and (degenerate) normal distributions, as important boundary cases with unbounded score functions. Under a reparameterization directly linking $m_1$ and $m_2$ to new parameters $p$ and $q,$ one can achieve the same generality, but with the added benefit of the standard normal distribution occurring as a limiting case when both $m_1$ and $m_2$ diverge to $\infty$ \citep{KalbfleischPrentice2e}.

Let $\mathcal{F}_{m_1,m_2}(\cdot)$ denote the CDF of a $\log F_{2m_1,2 m_2}$ random variable, and let $f_0(\cdot)$ denote its corresponding PDF  \citep[][eqn.\ (2.5)]{KalbfleischPrentice2e}. Then, using \eqref{eq:choice.for.a} with $f_0(\cdot)$ as specified, the score function 
reduces to \citep{McKean-Sievers-89}
\begin{eqnarray}\label{eq:gen.F.a}
a(u) &=& \frac{m_1 m_2 (\exp\{\mathcal{F}^{-1}_{m_1,m_2}(u)\} - 1)}{m_2+m_1 \exp\{\mathcal{F}^{-1}_{m_1,m_2}(u)\} }.
\end{eqnarray}
It can be shown that $-m_1 \leq a(u) \leq m_2$; hence, the results of the previous section apply to this choice of $a(\cdot)$ for finite values of $m_1$ and $m_2.$ If we define $F^{-1}_{2 m_1,2 m_2}(u)$ as the quantile function of the standard $F$ distribution with $2m_1$ and $2 m_2$ degrees of freedom, the above can be simplified; in particular we can just replace $\exp\{\mathcal{F}^{-1}_{m_1,m_2}(u)\}$ 
by $F^{-1}_{2 m_1,2 m_2}(u).$
From \eqref{eq:A.closed.form}, the function $A(u)$ corresponding to \eqref{eq:gen.F.a} is 
\begin{equation}
    A(u)= -\frac{m_1^{m_1}m_2^{m_2}}{B(m_1,m_2)}\frac{[F^{-1}_{2m_1,2m_2}(u)]^{m_1}}{[m_2+m_1F^{-1}_{2m_1,2m_2}(u)]^{m_1+m_2}}
\end{equation}
where $B(m_1,m_2)$ is the beta function. Fast numerical evaluation of $F^{-1}_{m_1,m_2}(\cdot),$ needed to calculate $a(u)$ and $A(u),$ is available in R and many other languages.

\section{Simulation Studies}
\label{sec:sim}
 We conducted simulation studies with the following goals.  First, we want to establish that parameter estimates and their variance-covariance estimates, as well as quasi-score and Wald tests, perform well in finite samples.  Second, we want to investigate how the choice of transformation $a(r)$ affects the performance of our estimators and tests.  Finally, we want to compare our results to those obtained by the Fygenson and Ritov approach described in section \ref{sec:R-est}.  
 To achieve these goals, we generated data under model \eqref{linear.model} in which 
 the distribution of $\epsilon$ followed a centered extreme value (Gumbel) distribution having mean zero and scale parameter $k=\frac{1}{2};$ that is, the resulting hazard function is $\lambda_F(u) = k \exp(k u - \xi),$ where  $\xi = 0.57721 \cdots$ is Euler's constant. 
 This choice corresponds to Weibull-distributed failure times with shape parameter $k=\frac{1}{2}$ and scale parameter 
 $\exp\{ 2 \xi + X^T_i\beta_0 \};$ equivalently, the
 conditional distribution $T^*|X$ is characterized
 by the hazard function $\lambda_{T^*}(t;X) = k \exp\{ - (\xi+ k X^\top \beta_0) \} t^{k-1}.$
 For our simulations, we considered data with $n=200$ observations and two covariates $X_1$ and $X_2.$ We generated $X_2 \sim \mbox{Bernoulli}(0.5)$ and, to create some correlation between $X_1$ and $X_2,$ generated $X_1 \sim Z + \frac{1}{2}X_2$ for $Z \sim N(0,1)$ independently of $X_2.$ 
Censoring times were distributed as $\log(C_i) \sim N(\mu_c=1.5,\sigma_c=2)$, resulting in an average censoring rate of approximately 34\%. 
For simplicity, we considered a range of $\beta_0$ values defined through the relationship $\beta_0= b \times (1,-1)^\top$ 
where $b$ varies in the interval $[-1,1].$

We considered 5 candidate functions $a(u)$ in our analyses:  $a(u)=u$ corresponding to the use of the Wilcoxon score, or equivalently `unweighted' R-estimators, of Section \ref{sec:R-est} (denoted raft.NoW); $a(u)=-1-\log(1-\frac{n}{n+1}u)$ where $n$ is the sample size, corresponding to the (shifted) optimal choice for Gumbel data (denoted raft.WW); and, $a(u)$ given in \eqref{eq:gen.F.a}, respectvely for $(m_1,m_2)=(1,10)$ (denoted raft.FW1), $(m_1,m_2)=(10,1)$ (denoted raft.FW2), and $(m_1,m_2)=(3,3)$ (denoted raft.FW3).  In the generalized $F$ family, the Gumbel distribution corresponds to $m_1=1, m_2 \rightarrow\infty;$ hence, we can anticipate $(m_1,m_2)=(1,10)$ to perform well and $(m_1,m_2)=(10,1)$ to perform comparatively less well (i.e., in terms of efficiency). For comparison, we also include results obtained using the Fygenson-Ritov estimator (denoted fraft) described in Section \ref{sec:R-est}.

In Table \ref{Table1}, the results summarized show that the bias of our proposed R-estimates is small, and that the estimated variances for 
\eqref{sig-hat-WLR} and \eqref{sig-hat-WLR.gen}, and for the parameter estimates as given in \eqref{eq:Omega.hat.huang}, perform well
for all five choices of $a(u)$. Importantly, the scale of $\Sigma$ depends on the choice of rank transformation $a(u)$ whereas the scale of $\Omega$ does not.  Relative efficiencies across different choices of $a(\cdot)$ can therefore be compared using their respective empirical estimates of $\Omega$.  For example, compared to the optimal weight function (i.e., as given in \eqref{eq:choice.for.a} when $F_0$ is the CDF of $\epsilon$), the use of the Wilcoxon score results in a relative efficiency of $1.14=1.506/1.322$ for parameter $\beta_1$ when $\beta_0=(0,0)$.  Note that the Fygenson and Ritov estimator has relative efficiency $1.28=1.689/1.322$ for $\beta_1$ compared with the optimal choice, and so performs worse than the Wilcoxon score. In this example, the generalized $F$ estimator with $(m_1,m_2)=(1,10)$ has relative efficiency of approximately 1.0 (i.e., compared to the optimal weight choice); this is consistent with the idea that $m_2=10$ is already large enough to approximate the relevant limiting result (i.e., while still maintaining a bounded score function).

\begin{table}
\scriptsize
\begin{tabular}[t]{|l|l|r|r|r|r|r|r|r|r|r|}
\hline
& &\multicolumn{2}{c|}{$Emp.\,var\,of\,$}&\multicolumn{2}{c|}{$Mean\,est\,%
var\,of\,$}&\multicolumn{1}{c|}{Bias}&\multicolumn{2}{c|}{$Emp.\, var\,of$}&\multicolumn{2}{c|}{$Mean\, est\,
var\,of$}\\
&  &\multicolumn{2}{c}{$\Psi \ (\times10^{-4})$}&\multicolumn{2}{|c|}{$\Psi \ (\times10^{-4})$}&\multicolumn{1}{c|}{$(\times10^{-2})$}&\multicolumn{2}{c|}{$\widehat{\beta} \ (\times10^{-1})$}&\multicolumn{2}{c|}{$\widehat{\beta}\ (\times10^{-1})$}\\
\hline
$\beta$&Method&$\Sigma_{11}$&$\Sigma_{22}$&$\widehat\Sigma_{11}$&$%
\widehat\Sigma_{22}$&$\overline{\beta}-\beta$&$\Omega_{11}$&$%
\Omega_{22}$&$\widehat\Omega_{11}$&$\widehat\Omega_{22}$\\
\hline
$(0,0)$&raft.NoW& 0.923 & 3.916 & 0.894 & 3.893 & (1.419,-0.794) & 1.506 & 0.381 & 1.571 & 0.377 \\ 
\hline
$(0,0)$&raft.WW& 7.472 & 31.686 & 7.331 & 30.678 & (1.268,-0.459) & 1.322 & 0.318 & 1.368 & 0.329 \\
\hline
$(0,0)$&raft.WF1& 7.350 & 31.170 & 7.205 & 30.197 & (1.258,-0.490) & 1.324 & 0.319 & 1.371 & 0.330 \\ 
\hline
$(0,0)$&raft.WF2& 10.011 & 42.538 & 9.898 & 45.811 & (2.700,-1.109) & 2.290 & 0.610 & 2.322 & 0.556 \\ 
\hline
$(0,0)$&raft.WF3& 14.043 & 59.617 & 13.765 & 60.051 & (1.730,-0.793) & 1.527 & 0.386 & 1.570 & 0.376 \\
\hline
$(0,0)$&fraft& 3.115 & 13.226 & 3.018 & 13.436 & (1.535,-0.959) & 1.689 & 0.437 & 1.756 & 0.423 \\
\hline
$(1,-1)$&raft.NoW& 0.927 & 3.755 & 0.914 & 3.962 & (2.027,-1.479) & 1.559 & 0.426 & 1.630 & 0.413 \\ 
\hline
$(1,-1)$&raft.WW& 7.384 & 28.941 & 7.409 & 29.785 & (1.777,-1.094) & 1.378 & 0.361 & 1.413 & 0.359 \\ 
\hline
$(1,-1)$&raft.WF1& 7.274 & 28.596 & 7.288 & 29.501 & (1.795,-1.133) & 1.376 & 0.362 & 1.420 & 0.360 \\
\hline
$(1,-1)$&raft.WF2& 10.038 & 41.745 & 9.757 & 43.881 & (3.358,-1.818) & 2.360 & 0.661 & 2.414 & 0.607 \\ 
\hline
$(1,-1)$&raft.WF3& 14.071 & 57.148 & 13.882 & 60.344 & (2.340,-1.473) & 1.577 & 0.431 & 1.626 & 0.412 \\
\hline
$(1,-1)$&fraft& 3.174 & 12.968 & 3.135 & 13.985 & (2.228,-1.709) & 1.759 & 0.488 & 1.835 & 0.465 \\ 
\hline
$(-1,1)$&raft.NoW& 0.927 & 3.768 & 0.913 & 3.846 & (1.227,-0.572) & 1.608 & 0.432 & 1.623 & 0.411 \\
\hline
$(-1,1)$&raft.WW& 7.385 & 29.106 & 7.379 & 29.144 & (1.277,-0.387) & 1.425 & 0.365 & 1.406 & 0.360 \\ 
\hline
$(-1,1)$&raft.WF1& 7.276 & 28.754 & 7.259 & 28.805 & (1.258,-0.405) & 1.428 & 0.367 & 1.410 & 0.360 \\ 
\hline
$(-1,1)$&raft.WF2 & 10.039 & 41.823 & 10.040 & 45.101 & (2.473,-0.730) & 2.395 & 0.675 & 2.395 & 0.601 \\
\hline
$(-1,1)$&raft.WF3 & 14.074 & 57.339 & 13.980 & 59.144 & (1.484,-0.577) & 1.627 & 0.441 & 1.618 & 0.410 \\
\hline
$(-1,1)$&fraft & 3.175 & 13.011 & 3.104 & 13.546 & (1.309,-0.620) & 1.795 & 0.493 & 1.822 & 0.460 \\
\hline
\end{tabular}
\caption{Simulations with Extreme Value Error. Sample size 200. Results based on 1,000 replicates.}
\label{Table1}
\end{table}

\begin{figure}[hbt]
{\centering\includegraphics[width=1.0\textwidth]{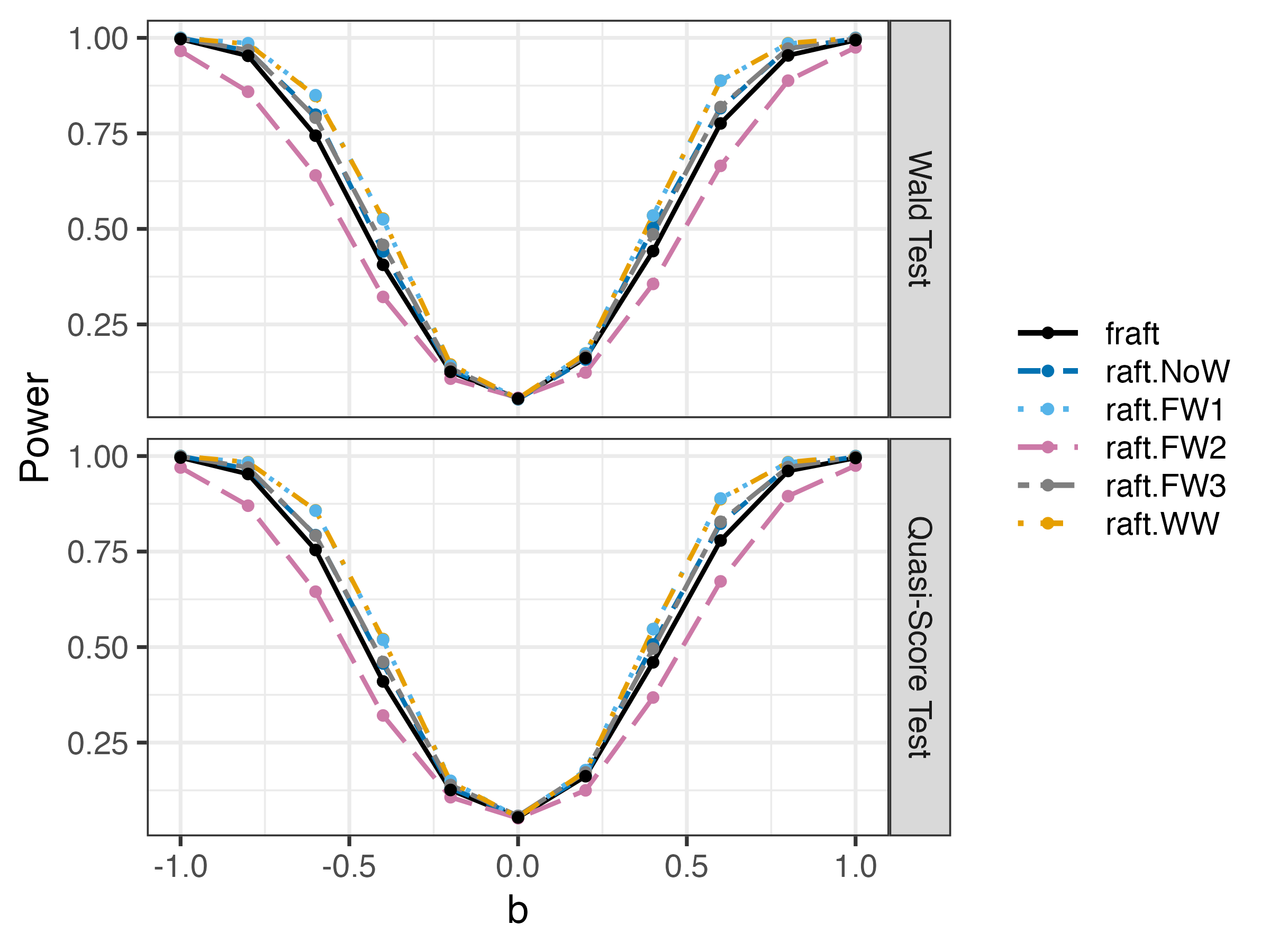}}
\caption{Power of the Wald and Quasi-Score tests to reject the hypothesis $\beta_0=(0,0)^\top$ for varying weight functions $a(u)$ and 
for the Fygenson and Ritov estimator (fraft) for alternatives of the form $\beta_0 = (b,-b)^\top,$ $b \in [-1,1].$}\label{fig:power.plot}
\end{figure}

To better illustrate the relative efficiencies of the estimators in our study, in Figure \ref{fig:power.plot} we also consider the power of both the Wald and quasi-score tests for each approach to reject the null hypothesis $H_0:\beta_0=(0,0).$
All choices of $a(\cdot)$, along with the Fygenson and Ritov estimator, result in tests with the appropriate size under $H_0.$ Figure \ref{fig:power.plot} also displays the estimated power of the indicated two-sided tests as $\beta_0 = (b,-b)^\top$ deviates from $H_0$ in the direction defined by $b \in [-1,1].$ The optimal weight choice and the generalized F score with $(m_1,m_2)=(1,10)$ are observed to have the highest power, and as might be expected, are very similar to each other. The powers achieved using the Wilcoxon score $a(u) = u,$ equal (i.e., up to location and scale) to a generalized F score with $(m_1,m_2)=(1,1),$ and the generalized F score with $(m_1,m_2)=(3,3)$, are next; here, both scores correspond to distributions of $\epsilon$ that are symmetric. Finally, the Fygenson and Ritov approach has the second-lowest power of all the methods considered here, followed by the poorly chosen generalized F score with $(m_1,m_2)=(10,1).$ 

\begin{figure}[htbp]
{\centering\includegraphics[width=1.0\textwidth]{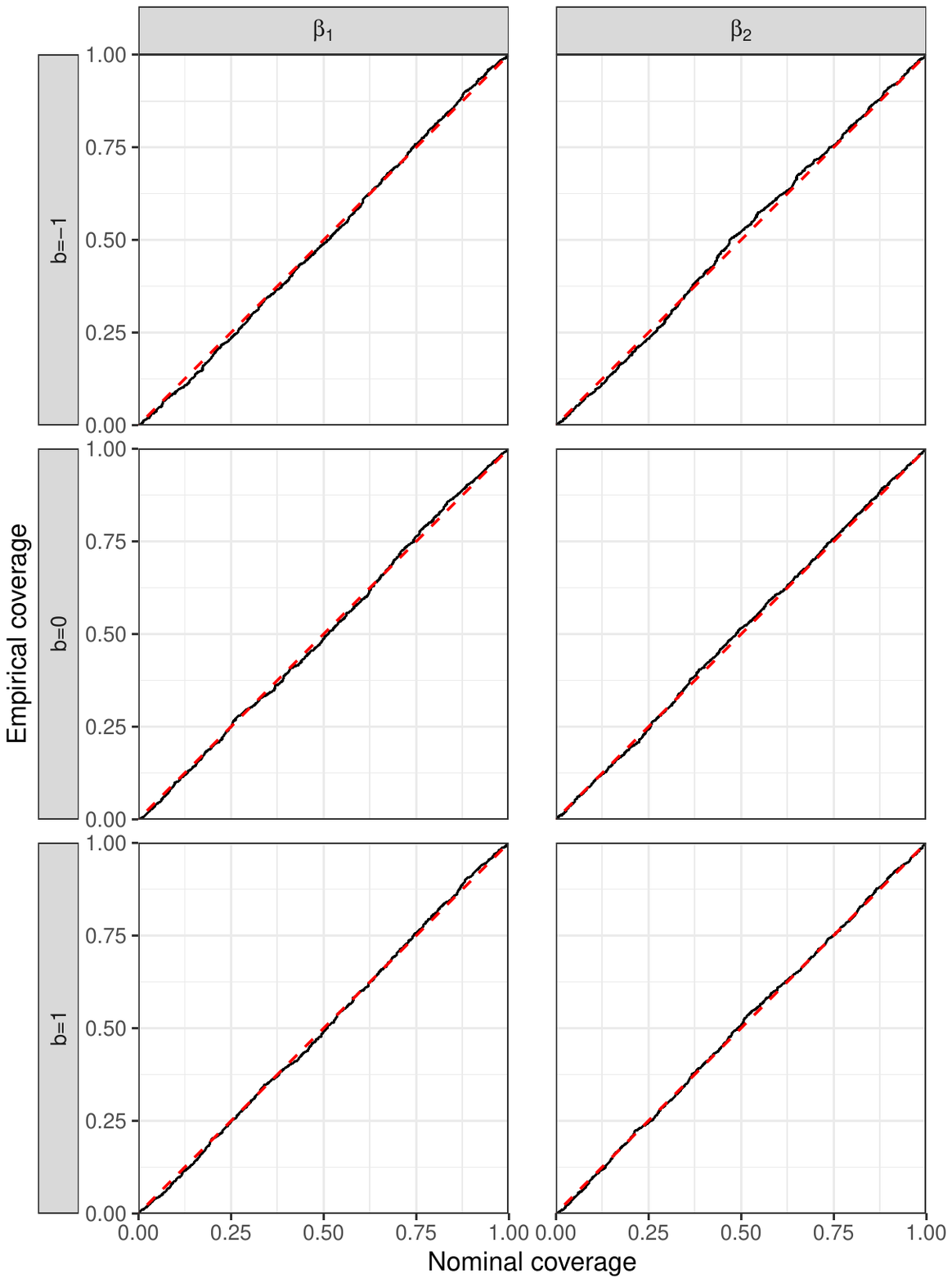}}
\caption{Empirical versus nominal coverage of (marginal) Wald-type confidence intervals for $\beta_1$ (1st column) and $\beta_2$ (2nd column) when 
$\beta_0 = (b,-b)^\top,$ $b \in [-1,1],$ when using the optimal weight function $a(u)=-1-\log(1-\frac{200}{201}u)$.  Red dashed line corresponds to equality of nominal and emprical coverage rates. }\label{fig:coverage.ww}
\end{figure}

\begin{figure}[htbp]
{\centering\includegraphics[width=1.0\textwidth]{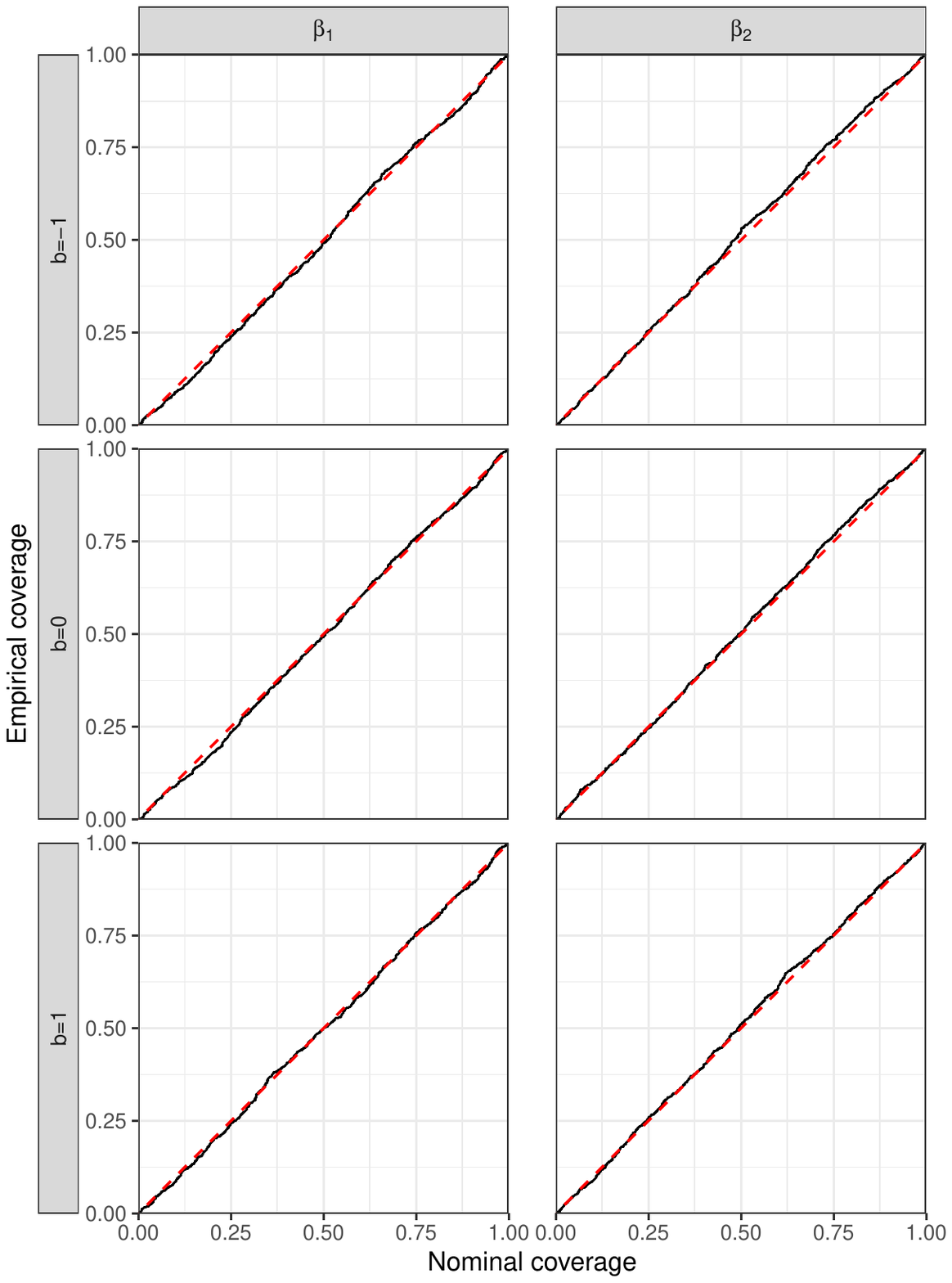}}
\caption{Empirical versus nominal coverage of (marginal) Wald-type confidence intervals for $\beta_1$ (1st column) and $\beta_2$ (2nd column) 
when $\beta_0 = (b,-b)^\top,$ $b \in [-1,1],$ using generalized $F$ weight function with $(m_1,m_2)=(10,1)$.  
Red dashed line corresponds to equality of nominal and emprical coverage rates.} \label{fig:coverage.f10.1}
\end{figure}

To confirm the asymptotic normality of the distribution of $\sqrt{n}(\widehat\beta-\beta_0)$ for $\beta_0 = (\beta_1,\beta_2)^\top$ and further illustrate the accuracy of the estimated variance $\widehat\Omega$, we also compared the empirical and nominal coverage of (marginal) confidence intervals for $\beta_j, j=1,2.$ Figure \ref{fig:coverage.ww} gives these plots for the optimal choice of $a(u)$ and Figure \ref{fig:coverage.f10.1} is for the worst-performing $(m_1,m_2)=(10,1)$.  The close agreement between empirical and nominal coverage suggests our R-estimators perform well in analyses of censored data from linear models across a range of choices for $a(\cdot).$

\clearpage

\section{Discussion}
\label{sec:discuss}

R-estimators of linear regression parameters have long been considered as an attractive alternative to ordinary least squares estimators.  In this paper, we have proposed a natural generalization of the notion of rank, and corresponding natural extensions of the R-estimating equation \eqref{R.est} for handling right-censored data in the case of linear and nonlinear score functions $a(\cdot).$  Our simulation results show that the proposed estimators perform well with Weibull-distributed failure times across a range of scores $a(\cdot),$ but that also that the choice of $a(\cdot)$ can affect the power and efficiency of the resulting estimators. As expected, the optimal choice of weight function led to the best performance; the use of a weight function that only approximates this optimal weight led to nearly identical performance. Our results further 
show that the use of the Wilcoxon score, which results in a weighted logrank statistic with weight given
by the Kaplan Meier (i.e., self-consistent)  estimator of the residual distribution, leads to higher power and efficiency when compared to the estimator of \cite{FygensonRitov94}, which corresponds to the Gehan-weighted logrank statistic with weight given
by $\sum_{j=1}^n Y_j(u;\beta).$ 

The paper initially focuses on the case of a linear-in-ranks score function, that is, $a(u) = u.$ One useful consequence of our formulation of the rank through \eqref{est.rank.bj} stems from Theorem \ref{nice result 2}, where it is shown that the sum of the ranks is constant regardless of $\beta,$ just as in the complete data setting. We have also shown that there is an exact relationship between the corresponding rank estimating function \eqref{R.est.cens} and the class of estimators respectively proposed in \citet{ritov1990} and \citet{tsiatis1990} for right-censored outcomes. 
Each of these results relies directly on the  self-consistent estimator for $S$ (hence $F$) and corresponding mid-CDF in the definition \eqref{est.rank.bj}; these non-asymptotic results hold for a general $F.$  Asymptotically, of course, many other estimators of $S$ ($F$) are possible, and it may be of interest to see whether other valid choices substantively influence finite sample performance.

To the author's knowledge, the developments in Section \ref{sec:genR} for a nonlinear score are new, particularly so the exact correspondence between \eqref{eq:gen-Wilcoxon-2} and \eqref{eq:gen-Wilcoxon-1} (i.e., through \eqref{eq:key-integral}), and the associated estimating equation \eqref{R.est.cens.gen}. These results are valid for general $F,$ and directly generalize those for a linear score in an interesting way. Notably, these developments do not specifically rely on the use of the mid-CDF $\widehat F^{\star}_{\beta}(t),$ but rather on the relationship specified in \eqref{eq:Fhat-a}, a particular generalization that (i) depends on $a(\cdot)$ and (ii) is designed to ensure that \eqref{eq:key-integral} holds. We further show how to extend these results to any proper CDF $H(\cdot);$ this is accomplished by introducing \eqref{eq:gamma_a}, defined for each $t$ as the expectation of a nonlinear function of a uniformly weighted average of $H(t)$ and $H(t-).$ This definition of $\gamma_a(t;H)$ reproduces \eqref{eq:Fhat-a} when $H(t) = \widehat F_{\beta}(t);$ more generally, it's use guarantees mathematical equivalence between $\eqref{eq:Gamma_a_1}$ and $\eqref{eq:Gamma_a_2},$ which generalizes \eqref{eq:key-integral}. 

A novel contribution of this paper is the demonstration of the exact interconnections between the generalization of the R-estimator for right-censored data and the class of weighted logrank statistics; in particular, and respectively, between \eqref{R.est.cens} and \eqref{WLR-Glen} and between \eqref{R.est.cens.gen} and a weighted logrank statistic that uses the weight function \eqref{eq:exact-weight}. Of course, as a consequence of these interconnections, the resulting general class of regression parameter estimators, hence methods for inference as currently outlined in Section \ref{sec:variance}, are not exactly new, particularly when $a(u) = u.$  One feature of these equivalence results is that they can be used motivate the use of specific weight functions that are not commonly used when constructing weighted log-rank statistics. For example, the literature has often focused on weights that are efficient and/or powerful for data that closely follow a Cox proportional hazards model. The class of weight functions considered in this paper instead correspond to optimal choices for R-estimators for data that follow an accelerated failure time (AFT) model with a given error distribution (equivalently, a linear model for the log-failure time), extended to account for the presence of right censoring. The relatively simple form of our censored-data R-estimators will also enable extending the proposed methods to more complicated data structures, such as clustered failure time data with right-censored outcomes; see, e.g., \citet{rlme2013} and references therein.

In this work, and consistent with the literature on R-estimators for uncensored data, we have focused on right-censored
outcomes and settings in which $a(\cdot)$ is presumed to be monotone. In the case of uncensored data, the monotonicity of $a(\cdot)$ in \eqref{R.est} leads to an estimator that is also the minimizer of a convex objective function in $\beta$, that is, the Jaeckel dispersion \citep{jaeckel1972estimating}. However, with right-censored data, monotonicity of \eqref{R.est.cens.gen} in $\beta$ cannot be expected even when $a(u) = u;$ hence, our proposed estimators already fail to be minimizers of some convex objective function. Moreover, the optimal score choice \eqref{eq:choice.for.a} (i.e., when $F_0(\cdot)$ is the CDF of $\epsilon$) is itself not guaranteed to be monotone.  For example, it can be shown that \eqref{eq:choice.for.a} implies $a(u)=\sin(2\pi u)$ when $\epsilon$ has a Cauchy distribution. These observations, and the continued applicability of results like those summarized in Sections \ref{sec:variance} and \ref{sec:nonlinear score} in the absence of censoring, therefore expand the class of novel rank estimators available in uncensored data problems to weighted R-estimators with non-monotone scores. Preliminary simulations using the Cauchy distribution and its optimal weight function (results not shown) suggest this may be a useful approach for both censored and uncensored data.

\section*{Acknowledgements}

This work was partially supported by the National Institutes of 
Health grant R01ES034021 (RLS) and R01GM147162  (NZ, GS).

\bibliographystyle{abbrvnat}
\bibliography{bibliography}

\newpage
\appendix
\section{Appendix}

We begin with a helpful result concerning the behavior of the mid-CDF. For any general proper CDF $H(t),$ define $H^\star(t) = \frac{1}{2}( H(t) + H(t-) )$ to be its
corresponding mid-CDF. Then:
\begin{equation}
\label{F-int}
\int_a^b H^\star(t) dH(t) = \frac{1}{2} \left[ H^2(b)  - H^2(a)  \right];
\end{equation}
see, for example, Ross (Elementary Analysis: Theory of Calculus, 2nd ed, Thm.\ 35.19). 
An immediate consequence of thethe integration-by-parts formula 
\eqref{F-int} is that
\begin{equation}
\label{S-int}
 \frac{1}{S(t)} \int_t^{\infty} F^\star(u)  d F(u) = 1 - \frac{S(t)}{2},
\end{equation}
where $S = 1-F.$

In some of the proofs, we will also make use of counting process notation, along with properties of
the self-consistent estimator of $\widehat S_{\beta}(t)$ for any fixed, finite $\beta.$ 
For $i=1,\ldots,n,$ recall that $N_i(u;\beta) = \Delta_i I\{ \tilde E_{i,\beta} \leq u\}$
and $Y_i(u;\beta) = I\{ \tilde E_{i,\beta} \geq u\};$ let $N_{+}(u;\beta) = \sum_{i=1}^n N_{i}(u,\beta)$ and
$Y_{+}(u;\beta) = \sum_{i=1}^n Y_{i}(u,\beta),$ and define $\widehat \Lambda_{\beta}(t)$ as
the Nelson-Aalen estimator corresponding to $\widehat S_{\beta}(t);$ see \eqref{eq:Lam-hat}. With
$\widehat F_{\beta}(t) = 1-\widehat S_{\beta}(t),$ we can also write
$d \widehat F_{\beta}(t) = \widehat S_{\beta}(t-) d \widehat \Lambda_{\beta}(t)$
and $\int_{-\infty}^{\infty} d \widehat F_{\beta}(u) = 1.$ 

We remark here that all such quantities assume, unless otherwise stated, the use of a modified dataset 
in which all observation times $T_i$ tied with $T_{(n)}$ have failure status $\Delta_i = 1;$ this is only done to 
preserve compatibility with the use of the self-consistent estimator $\widehat F_{\beta},$ and is
particularly salient for the rank-sum proof of Theorem \ref{really nice result},
as well as Theorem \ref{thm:Psi.eq.WLR} due to the appearance of $\widehat \Lambda_{\beta}(\cdot).$

\subsection{Proof of Theorem
\ref{nice result}}
\label{app: thm1proof}

{\sc Proof:}
The simplified formula \eqref{eq:simple-R} for $\mathcal{R}_{\beta_0}$ follows from the integral forms \eqref{F-int} and \eqref{S-int}. 
It is possible to prove the stated results for its mean and variance by making extensive use of integration-by-parts; below, we will instead proceed by making use of the results of  \citet{shepherd2016}, resulting in a much simpler and shorter proof.

In particular, it is easy to show that \eqref{eq:simple-R} can be written
\[
\mathcal{R}_{\beta_0} = 
\frac{1}{2} + \frac{1}{2} 
r(\tilde E_{\beta_0}, F, \Delta)
\]
where $r(y,F,\Delta) = F(y) - \Delta (1-F(y-))$ is the probability-scale residual for right-censored data introduced in \citet{shepherd2016} for any underlying failure CDF $F$.  Results given in \citet[Sec.\ 5]{shepherd2016} now imply
$E[ r(\tilde E_{\beta_0}, F, \Delta) | X] = 0$ provided
that $T^* \bot C^* \mid X;$ hence,  it follows immediately that
$E(\mathcal{R}_{\beta_0} | X) = E(\mathcal{R}_{\beta_0}) = 1/2.$ Moreover, in the case where the
underlying distribution $F$ is continuous,  their results  additionally imply
\[
\mbox{Var}(\mathcal{R}_{\beta_0} | X) = 
\frac{1}{12} \left( 1 - E( S^3(E_{\beta_0}) | X) \right);
\] 
a corresponding version of this formula is also given with a discreteness correction
that additionally involves the cubes of the jump sizes in the case where $F$ has jumps.  
Because $E(\mathcal{R}_{\beta_0} | X) = 1/2,$ we also have $\mbox{Var}( E(\mathcal{R}_{\beta_0} | X)  ) = 0$ and hence that
\[
\mbox{Var}(\mathcal{R}_{\beta_0} ) = E\left( \mbox{Var}(\mathcal{R}_{\beta_0} | X)  \right) = \frac{1}{12} \left( 1 - E( S^3(E_{\beta_0}) ) \right),
\]
as was to be proved.

\subsection{Proof of Theorem
\ref{thm:Psi.eq.WLR}}
\label{app: Psi.eq.WLR}

Define $
M(u) = 1 -  \widehat{S}_{\beta}(u) / 2;
$
then, after some algebraic manipulation,
observe that \eqref{R.est.cens}
can be rewritten
\begin{align*}
\Psi_n(\beta) & =  \sum_{i=1}^n  ( X_{i} - \bar X)
\left[ \Delta_i \widehat F^{\star}_{\beta}( \tilde E_{i,\beta} )
+ (1-\Delta_i) M( \tilde E_{i,\beta} ) \right] \\
& = 
\sum_{i=1}^n \int_{-\infty}^{\infty}
\widehat F^{\star}_{\beta}(u) (X_i - \bar X(u;\beta) )
d N_i(u;\beta)
+ \int_{-\infty}^{\infty}
\widehat F^{\star}_{\beta}(u) (\bar X(u;\beta) - \bar X)
d N_{+}(u;\beta) \\
&\hspace{5mm}
+ \sum_{i=1}^n ( X_{i} - \bar X) M( \tilde E_{i,\beta} )
- \sum_{i=1}^n (X_i- \bar X) \int_{-\infty}^{\infty}
M(u) d N_i(u;\beta).
\end{align*}
Letting $W(u) = \widehat F^{\star}_{\beta}(u) 
 - M(u),$ further algebra shows that
\begin{align}
\nonumber
\Psi_n(\beta) & = 
\sum_{i=1}^n \int_{-\infty}^{\infty}
W(u) (X_i - \bar X(u;\beta) )
d N_i(u;\beta)
 \\
\label{eq: key psi rep}
&\hspace{5mm}
+ \int_{-\infty}^{\infty}
W(u) (\bar X(u;\beta) - \bar X)
d N_{+}(u;\beta)
+ \sum_{i=1}^n ( X_{i} - \bar X) M( \tilde E_{i,\beta} ).
\end{align}
Also, observe that
\begin{eqnarray*}
\sum_{i=1}^n ( X_{i} - \bar X) M( \tilde E_{i,\beta} )
& = & - \frac{1}{2} \sum_{i=1}^n ( X_i - \bar X)
\widehat{S}_{\beta}(\tilde E_{i,\beta}) \\
& = &
- \sum_{i=1}^n (X_i - \bar X) 
 \Bigl( \frac{\widehat{S}_{\beta}(\tilde E_{i,\beta})}{2}
 - \frac{1}{2} \Bigr).
\end{eqnarray*}
Because $\widehat{S}_{\beta}(u-)$ is the self-consistent
estimator of $S,$ we have
\[
\int_{-\infty}^{\tilde E_{i,\beta}}
\widehat{S}_{\beta}(u-) \frac{d N_+(u;\beta)}{Y_+(u;\beta)}
= \int_{-\infty}^{\tilde E_{i,\beta}} d 
\widehat{F}_{\beta}(u) =
\widehat{F}_{\beta}(\tilde E_{i,\beta})
= 1 - \widehat{S}_{\beta}(\tilde E_{i,\beta})
= - 2 \left(\frac{\widehat{S}_{\beta}(\tilde E_{i,\beta})}{2}
 - \frac{1}{2}. 
\right)
\]
hence, 
\begin{align*}
\sum_{i=1}^n & ( X_{i} - \bar X ) M( \tilde E_{i,\beta} )
 = 
 \frac{1}{2} \sum_{i=1}^n ( X_i - \bar X) 
\int_{-\infty}^{\tilde E_{i,\beta}}
\widehat{S}_{\beta}(u-) \frac{d N_+(u;\beta)}{Y_+(u;\beta)}\\
 & = 
 \frac{1}{2} \left\{  \sum_{i=1}^n X_i \int_{-\infty}^{\infty} 
\widehat{S}_{\beta}(u-) Y_i(u;\beta) \frac{d N_+(u;\beta)}{Y_+(u;\beta)} 
-\bar X \sum_{i=1}^n \int_{-\infty}^{\infty}
\widehat{S}_{\beta}(u-) Y_i(u;\beta) \frac{d N_+(u;\beta)}{Y_+(u;\beta)} 
\right\} \\
 & = 
  - \int_{-\infty}^{\infty} 
 -\frac{\widehat{S}_{\beta}(u-) }{2} \bar X(u;\beta) d N_+(u;\beta)
+ \bar X  \int_{-\infty}^{\infty}
-\frac{\widehat{S}_{\beta}(u-) }{2} d N_+(u;\beta)
 \\
 & = 
  -\int_{-\infty}^{\infty} W(u)
   \bar X(u;\beta) d N_+(u;\beta)
+ \bar X  \int_{-\infty}^{\infty}
W(u) d N_+(u;\beta),
\end{align*}
the last step following from the fact that
\begin{eqnarray}
\label{eq:W-res}
W(u)= \widehat F^{\star}_{\beta}(u) 
 - M(u)  =  -\frac{1}{2} \hat S_{\beta}(u-).
\end{eqnarray}
Therefore, substituting the resulting expression
into \eqref{eq: key psi rep} and using
\eqref{eq:W-res},
\begin{align*}
\Psi_n(\beta) & = 
\sum_{i=1}^n \int_{-\infty}^{\infty}
W(u) (X_i - \bar X(u;\beta) )
d N_i(u;\beta)
+ \int_{-\infty}^{\infty}
W(u) (\bar X(u;\beta) - \bar X)
d N_{+}(u;\beta) \\
&\hspace{5mm}
- 
\int_{-\infty}^{\infty} W(u)
   \bar X(u;\beta) d N_+(u;\beta)
+ \bar X  \int_{-\infty}^{\infty}
W(u) d N_+(u;\beta) \\
& = -\frac{1}{2} \sum_{i=1}^n \int_{-\infty}^{\infty}
\widehat S_{\beta}(u-) (X_i - \bar X(u;\beta) )
d N_i(u;\beta);
\end{align*}
which proves the required result without making
use of the identities used in \citet{ritov1990}.

\subsection{Regularity
Conditions for Asymptotic
Results}
\label{app: asycond}
In \citet{Ying1993}, the following regularity conditions are imposed:
\begin{enumerate}
\item $(T^*_i,C^*_i,X_i), i = 1,\ldots,n$ are independent.
\item For $i = 1,\ldots,n,$ 
\begin{enumerate}
\item $\| \beta_0 \| < \infty$ and $P\{ \| X_i \| \leq M \} = 1$ for some given $M < \infty$ (i.e., covariates are uniformly bounded)
\item Given $X_i,$ $T^*_i \bot C^*_i$ (equivalently: $\epsilon_i \bot E_{\beta_0,i}$). 
\item $\epsilon_i | X_i \sim F,$ where $F$ has bounded first ($f$) and second ($f'$) derivatives, and
\[
\int \left( \frac{d}{ds} \log f(s) \right)^2 f(s) ds < \infty.
\]
\item The distribution of $\log C^*_i$ has CDF $G_i,$ where $G_i$ has a bounded PDF.
\end{enumerate}
\item $\sup_{i} E | \epsilon_i \wedge E_{\beta_0,i} |^c < \infty$ for some $c > 0.$
\end{enumerate}

In \citet{Jin2004}, Conditions 1 and
2(a) - 2(c) are imposed;
the restrictive conditions on censoring, particularly
2(d) but also Condition 3, are removed and replaced by the following:
\begin{itemize}
\item The eigenvalues of the empirical analog of $\Sigma,$ defined in \eqref{eq:sigma}, are bounded away from
zero for all $n$ sufficiently large.
\end{itemize}

The above conditions are sufficient
to establish the asymptotic behavior of \eqref{T-fun},
and the corresponding $\widehat \beta,$ when $w(u;\beta) = 1.$ For a more general weight, an additional condition governing
the behavior of $w(u;\beta)$ as a function
of $(u,\beta)$ is needed. For example, 
\citet[][Sec.\ 3, Condition\ 5]{Ying1993} imposes 
a 3-part condition to ensure sufficiently
good behavior, and notes that this sufficient condition
is satisfied by $\widehat S_{\beta}(\cdot).$
In both \citet{Huang2002} and \citet{Strawderman2005},
a different sufficient condition is imposed; in the current
setting, this condition requires (i) $w(u;\beta)$ to be
of bounded variation and (ii) almost sure convergence
of $w(u;\beta)$ to a function $w(u)$ that is 
uniformly continuous for $u \in (-\infty,\infty)$ and
$\beta$ in some open neighborhood of $\beta_0.$

\subsection{Proof of Theorem \ref{thm:new var}}
\label{app:new var proof}
\subsubsection{A decomposition of $\Sigma$}
Recall that
\[
\Theta_j(u) = \lim_{n \rightarrow \infty} \frac{1}{n} \sum_{i=1}^n E[  X_i^{\otimes j} Y_i(u,\beta_0) ].
\]
Under the i.i.d.\ and continuity assumption,
\[
\Theta_j(u)
= S(u) E\bigl[
X^{\otimes j} Y_0(u) \bigr],
\]
where $Y_0(u) = I\{ E_{\beta_0} \geq u \}$ and
$E_{\beta_0}$ is defined as in the
Section \ref{app: thm1proof}.
It follows that
\[
\frac{\Theta_j(u)}{\Theta_0(u)}
= \frac{E\bigl[
X^{\otimes j} Y_0(u) \bigr]}{
E\bigl[
Y_0(u) \bigr]
}
\]
Defining
\[
\mu_{X|E_{\beta_0}}(u) =
\frac{\Theta_1(u)}{\Theta_0(u)} = \frac{E\bigl[
X Y_0(u) \bigr]}{
E\bigl[
Y_0(u) \bigr]},
\]
we can then write
\[
{\mathcal H}(u) = \frac{\Theta_2(u)}{\Theta_0(u)} - \left(\frac{\Theta_1(u)}{\Theta_0(u)}\right)^{\otimes 2} = 
\frac{E\bigl[
Y_0(u)
(X - \mu_{X|E_{\beta_0}}(u))^{\otimes 2}
\bigr]}{E[Y_0(u)] }; 
\]
consequently,
${\mathcal H}(u) \Theta_0(u) = V(u) S(u)$ where
\[
{\mathcal V}(u) = E\bigl[
Y_0(u) 
(X - \mu_{X|E_{\beta_0}}(u))^{\otimes 2}
\bigr].
\]
It follows that
\eqref{eq:sigma}
can be rewritten
\begin{equation}
\label{eq:sigma2}
\Sigma = \frac{1}{4} \int_{-\infty}^{\infty} S^2(u) {\mathcal V}(u) d F(u).
\end{equation}
Straightforward algebra 
shows
\[
{\mathcal V}(u) =
E\bigl[
Y_0(u) (X - \mu_{X})^{\otimes 2}
\bigr]
-
(\mu_{X|E_{\beta_0}}(u) - \mu_{X})^{\otimes 2}
E\bigl[ Y_0(u) \bigr];
\]
consequently,
$\Sigma = \Sigma_1 - \Sigma_2$
where
\[
\Sigma_1 = \frac{1}{4} \int_{-\infty}^{\infty} 
E\bigl[
Y_0(u) (X - \mu_{X})^{\otimes 2}
\bigr]
S^2(u) d F(u)
\]
and
\[
\Sigma_2 = \frac{1}{4} \int_{-\infty}^{\infty} 
(\mu_{X|E_{\beta_0}}(u) - \mu_{X})^{\otimes 2}
E\bigl[
Y_0(u) \bigr]
S^2(u) d F(u).
\]

Since both $\Sigma_1$ and $\Sigma_2$ are non-negative definite, the matrix $\Sigma_1$ necessarily provides 
a conservative assessment of variability. In the next subsection, we show that $\Sigma_1$ can be simplified
and that it has an interesting and useful interpretation.

\subsubsection{A simplification of $\Sigma_1$}

Letting
$\Sigma_X = E[ (X - \mu_{X})^{\otimes 2}],$
we can write
\begin{align*}
\Sigma_1 & = \frac{\Sigma_X}{4} \int_{-\infty}^{\infty} 
S^2(u) d F(u)
-
\frac{1}{4} \int_{-\infty}^{\infty} 
E\bigl[
(1-Y_0(u)) (X - \mu_{X})^{\otimes 2}
\bigr]
S^2(u) d F(u) \\
& = \frac{\Sigma_X }{12} 
-
\frac{1}{4} \int_{-\infty}^{\infty} 
E\bigl[
(1-Y_0(u)) (X - \mu_{X})^{\otimes 2}
\bigr]
S^2(u) d F(u),
\end{align*}
the second result following from the
fact that $\int_{-\infty}^{\infty} 
S^2(u) d F(u) = 1/3.$ Recall from
Section \ref{app: thm1proof} that 
$G(u|X) = P\{ E_{\beta_0} \leq \infty \mid X \};$
then, because
$
3 S^2(u) dF(u) = -d S^3(u),
$
\begin{align*}
\frac{1}{4} \int_{-\infty}^{\infty} &
E\bigl[ I\{ E_{\beta_0} < u \} (X - \mu_{X})^{\otimes 2}
\bigr]
S^2(u) d F(u) \\
& = 
- \frac{1}{12} E\biggl[ (X - \mu_{X})^{\otimes 2} 
\biggl\{ \int_{-\infty}^{\infty} 
G(u-|X)
d [S^3(u)] \biggr\} \biggr] \\
& = 
- \frac{1}{12} E\biggl[ (X - \mu_{X})^{\otimes 2}
\biggl\{G(\infty \mid X)
S^3(\infty) - 
G(-\infty \mid X) S^3(-\infty)
- 
\int_{-\infty}^{\infty} 
S^3(u) d G(u \mid X)
\biggr\} \biggr] \\
& = \frac{1}{12} E\biggl[ (X - \mu_{X})^{\otimes 2}
\int_{-\infty}^{\infty} 
S^3(u) d G(u \mid X)\biggr] \\
& = \frac{1}{12} E\bigl[ (X - \mu_{X})^{\otimes 2}
S^3(E_{\beta_0}) \bigr];
\end{align*}
therefore, 
\begin{align*}
\Sigma_1 & =  \frac{\Sigma_X }{12} 
-
\frac{1}{12} E\bigl[ (X - \mu_{X})^{\otimes 2}
S^3(E_{\beta_0}) \bigr].
\end{align*}

To better understand the nature of this last expression, we now establish how $\Sigma_1$ is related to $\mbox{Var}(n^{-1/2} \Psi_n(\beta_0)),$ where $\Psi_n(\beta)$ is given in  \eqref{R.est.cens}.
Define $\mathcal{R}_{\beta_0,i} = \mathcal{R}(\tilde E_{i,\beta_0})$ as the population analog of \eqref{est.rank.3}; that is,
\begin{eqnarray*}
\mathcal{R}_{\beta_0,i} 
      = \Delta_i  F^{\star}( \tilde E_{i,\beta_0} ) + (1-\Delta_i)\left[ 1-\frac{1}{2} S( \tilde E_{i,\beta_0} ) \right].
\end{eqnarray*}
Also, let $\tilde \Psi_n(\beta_0) =  \sum_{i=1}^n
(X_i-\bar X) \mathcal{R}_{\beta_0,i}
=  \sum_{i=1}^n
(X_i-\bar X) (\mathcal{R}_{\beta_0,i} - 0.5).$
Then, under the imposed regularity conditions,
\begin{align*}
n^{-1/2} \tilde \Psi_n(\beta_0) & =
n^{-1/2} \sum_{i=1}^n (X_i-\mu_X) (\mathcal{R}_{\beta_0,i} - 0.5) 
- n^{-1/2} (\bar X - \mu_X) \sum_{i=1}^n (\mathcal{R}_{\beta_0,i} - 0.5) \\
& =
n^{-1/2} \sum_{i=1}^n (X_i-\mu_X) (\mathcal{R}_{\beta_0,i} - 0.5) 
- n^{1/2} (\bar X - \mu_X) \times \frac{1}{n} \sum_{i=1}^n (\mathcal{R}_{\beta_0,i} - 0.5) \\
&= n^{-1/2} \sum_{i=1}^n (X_i-\mu_X) (\mathcal{R}_{\beta_0,i} - 0.5) + o_P(n^{-1/2})
\end{align*}
and, as a result of the i.i.d.\ assumption, 
\begin{eqnarray*}
\mbox{Var}(n^{-1/2} \tilde \Psi_n(\beta_0) ) 
& \rightarrow & \mbox{Var}\{ (X_1-\mu_X) \{ \mathcal{R}_{\beta_0,1} - 0.5 \} \} 
\,=\,
E\left[ (X_1 - \mu_X)^{\otimes 2} \mbox{Var}( \mathcal{R}_{\beta_0,1} | X_1 ) \right], 
\end{eqnarray*}
where the second equality follows from the conditional variance formula since
\[
\mbox{Var}( E[ (X_1 - \mu_X)^{\otimes 2} \{ \mathcal{R}_{\beta_0,1} - 0.5 \} |X_1 ] ) = \mbox{Var}( (X_1 - \mu_X)^{\otimes 2}  E[ \mathcal{R}_{\beta_0,1} - 0.5 |X_1 ] ) = 0.\\
\]
Using the results from Theorem \ref{nice result}, 
\[
\mbox{Var}(\mathcal{R}_{\beta_0,1} | X_1) = \frac{1}{12} \left( 1 - E( S^3(E_{\beta_0,1} ) | X_1)  \right);
\]
it is then immediate that
\begin{align*}
\mbox{Var}(n^{-1/2} \tilde \Psi_n(\beta_0) ) 
& \rightarrow   E\Bigl[ (X_1 - \mu_X)^{\otimes 2} 
\mbox{Var}(\mathcal{R}_{\beta_0,1} | X_1) \Bigr]
= \Sigma_1
\end{align*}
as $n \rightarrow \infty.$
Consequently, $\Sigma_1$ represents the limiting variance of $n^{-1/2} \tilde \Psi_n(\beta_0).$ 
The product form of this variance is consistent with the observation that  $X_i$ and $R_{\beta_0,i}$ are uncorrelated; this occurs because 
\[
E[ n^{-1} \tilde \Psi_n(\beta_0) ] = 0
= E[ (X_1 - \mu_X) \{ \mathcal{R}_{\beta_0,1} - 0.5 \} ].
\]

Since $\Sigma = \Sigma_1 - \Sigma_2,$ one can therefore view the additional contribution of $\Sigma_2$ as stemming from the need to replace $\mathcal{R} ( \tilde E_{i,\beta_0} )$ by $\widehat{\mathcal{R}}( \tilde E_{i,\beta_0} )$ when estimating $\beta$ via \eqref{R.est.cens}, which actually reduces variability when compared to  using the (unknown) distribution function of the residuals.

\subsection{Proof of Theorem \ref{really nice result}}
\label{app: thm5proof}

Fix any finite $\beta$ and recall the definition of the processes
$N_i(u;\beta)$ $Y_i(u;\beta),$ and set
$M_i(u;\beta) = N_i(u;\beta) - \int_{-\infty}^u Y_i(s;\beta) d \Lambda_{H}(s),$ where $\Lambda_{H}$ is the cumulative
hazard function corresponding to some arbitrary proper CDF $H$. Also,
 let $\gamma_a(t;H)$ and $\Gamma_a(t;H)$ be defined as in \eqref{eq:gamma_a} and \eqref{eq:Gamma_a_1}, 
 respectively, and define for any suitable function $\gamma(\cdot)$ the functional
\begin{equation}
\label{eq:Ritov-wt}
W_H \gamma(t) = \gamma(t) - \frac{\int_t^{\infty} \gamma(s) d H(s)}{1-H(t)};
\end{equation}
then, for each $t$ such that $H(t) < 1$ and letting $d \Lambda_H(s) = d H(s) / (1-H(s-)),$
\begin{equation}
\label{eq:Ritov-equiv}
\int_{-\infty}^t W_H \gamma(s) d \Lambda_H(s) =  \int_{-\infty}^{\infty} \gamma(s) d H(s)
- \frac{\int_t^{\infty} \gamma(s) d H(s)}{1-H(t)}.
\end{equation}
This last identity is proved as part of Proposition 3.1 in \citet{ritov1990}, where it is additionally assumed 
that $\int_{-\infty}^{\infty} \gamma(s) d H(s) = 0.$ Then,
\begin{align} 
\nonumber
\int_{-\infty}^{\infty} & W_H \gamma_a(u;H) d M_i(u;\beta) 
= \int_{-\infty}^{\infty} W_H \gamma_a(u;H)  d N_i(u;\beta) - \int_{-\infty}^{\infty} W_H \gamma_a(u;H)  Y_i(u;\beta) d \Lambda_H(u) \\
\nonumber
& =  \int_{-\infty}^{\infty} W_H \gamma_a(u;H)  d N_i(u;\beta) - \int_{-\infty}^{\tilde E_{i,\beta}} W_H \gamma_a(u;H)  d \Lambda_H(u) \\
\label{eq:BJ-type-ident}
& =  \int_{-\infty}^{\infty} 
W_H \gamma_a(u;H)
d N_i(u;\beta) - 
\left( 
\int_{-\infty}^{\infty} \!\! \gamma_{a}(u;H) d H(u)
- \Gamma_a(\tilde E_{i,\beta})
\right),
\end{align} 
the last step following from first applying \eqref{eq:Ritov-equiv} and then using \eqref{eq:Gamma_a_1}. Using \eqref{eq:Ritov-wt},
\begin{align*}
\int_{-\infty}^{\infty}  
W_H \gamma_a(u;H)
d N_i(u;\beta) &
= \int_{-\infty}^{\infty} \left( \gamma_{a}(u;H) - \Gamma_a(u;H) \right) d N_i(u;\beta) \\
& = \Delta_i \left( \gamma_{a}(\tilde E_{i,\beta};H) - \Gamma_a(\tilde E_{i,\beta};H) \right); 
\end{align*}
it follows upon substitution into \eqref{eq:BJ-type-ident}, simplification and rearrangement of terms that
\begin{align}
\nonumber
\Delta_i  \gamma_a(  \tilde E_{i,\beta};H ) & + (1-\Delta_i) \Gamma_a( \tilde E_{i,\beta} ;H) \\
& = 
\label{eq:proof-ident}
\int_{-\infty}^{\infty} W_H \gamma_a(u;H) d M_i(u;\beta)  + \int_{-\infty}^{\infty}  \gamma_{a}(u;H) d H(u).
\end{align}
The identity \eqref{eq:proof-ident} is the key to proving both results stated in the theorem. \\

\noindent
\underline{Proof that $E(\mathcal{R}^{(a)}_{\beta_0} \mid X) = E(\mathcal{R}^{(a)}_{\beta_0}) = A(1) - A(0).$ }\\

Recall that
$
\mathcal{R}^{(a)}_{\beta_0} = \Delta \gamma_a(\tilde E_{\beta_0};F) + (1-\Delta) \Gamma_a(\tilde E_{\beta_0};F)
$
where $\tilde E_{\beta_0} = Y - X^T \beta_0$ is the true observed residual,
$\Delta$ is the true failure status,  and $F$ be 
the true CDF of $\epsilon = \log T^* - X^T \beta_0.$ Dropping all subject indices,
define $N(u) = N(u;\beta_0) = I\{ \tilde E_{\beta_0} \leq u, \Delta = 1 \},$ $Y(u) =  Y(u;\beta_0) = I\{ \tilde E_{\beta_0} \geq u \},$ and
$M(u) = N(u) - \int_{-\infty}^u Y(s) d \Lambda_F(s),$ where $\Lambda_F$ is the cumulative
hazard function corresponding to $F$. Then, using \eqref{eq:proof-ident},
\[
\mathcal{R}^{(a)}_{\beta_0}  = 
\int_{-\infty}^{\infty} W_H \gamma_a(u;F) d M(u)  
+ \int_{-\infty}^{\infty}  \gamma_{a}(u;F) d F(u).
\]
Under the conditions of this paper,
$M(u)$ is a mean-zero martingale; 
moreover, in view of \eqref{eq:Gamma_a_2},
\[
\int_{-\infty}^{\infty}  \gamma_{a}(u;F) d F(u) = A(1) - A(0)
\]
Thus, since $\gamma_a(\cdot;F)$ is a nonrandom function, hence predictable, 
\[
E\bigl[ \mathcal{R}^{(a)}_{\beta_0} |X ] = A(1) - A(0) = E\bigl[ \mathcal{R}^{(a)}_{\beta_0} ],
\]
proving the desired result.

\vspace*{0.1in}
\noindent \underline{Proof of \eqref{est.ranksum}}\\

Below, we prove that 
$\sum_{i=1}^n \widehat{\mathcal{R}}^{(a)} ( \tilde E_{i,\beta} ) =  n \int_{-\infty}^{\infty} \gamma_a(t; \widehat F_{\beta}) d \widehat F_{\beta}(t),$
where
\[
\widehat{\mathcal{R}}^{(a)}  ( \tilde E_{i,\beta} )
= \Delta_i  \gamma_a(\tilde E_{i,\beta}; \widehat F_{\beta}) + (1-\Delta_i) \Gamma_a( \tilde E_{i,\beta};\widehat F_{\beta} )
\]
and $\gamma_a(t;\widehat F_{\beta})$ and $\Gamma_a(t;\widehat F_{\beta})$ be defined as in \eqref{eq:gamma_a} and \eqref{eq:Gamma_a_1}
with $H = \widehat F_{\beta}.$ The stated result will then follow immediately from the fact that
\[
 \int_{-\infty}^{\infty} \gamma_a(t; \widehat F_{\beta}) d \widehat F_{\beta}(t)
= 
\sum_{-\infty < t < \infty}
\frac{A( \widehat  F_{\beta}(t))  - A( \widehat  F_{\beta}(t-) )}{\widehat  F_{\beta}(t)  - \widehat  F_{\beta}(t-)}
\times \Bigl(
\widehat  F_{\beta}(t)  - \widehat  F_{\beta}(t-)
\Bigr) 
= A(1) - A(0).
\]

To prove that $\sum_{i=1}^n \widehat{\mathcal{R}}^{(a)} ( \tilde E_{i,\beta} ) =  n \int_{-\infty}^{\infty} \gamma_a(t; \widehat F_{\beta}) d \widehat F_{\beta}(t),$
consider \eqref{eq:proof-ident} with $H =  \widehat  F_{\beta};$ then, we can write
\[
\widehat{\mathcal{R}}^{(a)}  ( \tilde E_{i,\beta} )
=
\int_{-\infty}^{\infty} W_{\widehat  F_{\beta}} \gamma_a(u;\widehat  F_{\beta}) d M_i(u;\widehat \beta)  + \int_{-\infty}^{\infty}  \gamma_{a}(u;\widehat  F_{\beta}) d \widehat  F_{\beta}(u).
\]
Consequently,
\[
\sum_{i=1}^n \widehat{\mathcal{R}}^{(a)}  ( \tilde E_{i,\beta} )
= \int_{-\infty}^{\infty} W_{\widehat  F_{\beta}} \gamma_a(u;\widehat  F_{\beta}) d M_+(u;\widehat \beta) 
+ n \int_{-\infty}^{\infty}  \gamma_{a}(u;\widehat  F_{\beta}) d \widehat  F_{\beta}(u). 
\]
However, in view of \eqref{eq:Lam-hat}, 
\[
d M_+(u;\widehat \beta) = d N_+(u;\widehat \beta) - Y_+(u;\widehat \beta) d \widehat \Lambda_{\widehat \beta}(u) =0,
\]
so $\sum_{i=1}^n \widehat{\mathcal{R}}^{(a)}  ( \tilde E_{i,\beta} ) = n \int_{-\infty}^{\infty}  \gamma_{a}(u;\widehat  F_{\beta}) d \widehat  F_{\beta}(u) = n (A(1) - A(0)),$ as desired.

\end{document}